\documentstyle[12pt]{article}
\title{Partial Differential Equations of Physics}
\author{Robert Geroch\\Enrico Fermi Institute, 5640 Ellis Ave,
Chicago, Il - 60637}
\begin{document}
\maketitle 

\section{Introduction}

The physical world is traditionally organized into various
systems:  electromagnetism, perfect fluids, Klein-Gordon
fields,
elastic media, gravitation, etc.  Our descriptions of these individual
systems have
certain features in common:  Use of fields on a fixed
space-time manifold $M$, a geometrical interpretation of
the fields in terms of $M$, partial differential
equations on these fields, an initial-value formulation for
 these equations.  Yet beyond these common features
there are numerous differences of detail:  Some systems of equations
are linear, and some are not; some have constraints, and some
do not; some arise from Lagrangians, and some do not; some
are first-order, and some higher-order. Systems also differ
in other respects, 
e.g., as to what fields they need as 
background, what interactions they permit (or require).  It
almost seems as though, in the end, every physical system has its
own special character. 

It might be useful to have a systematic treatment of the
fields and equations that arise in the description of physical
systems.  Thus, there would be a general definition of a 
``field", and a general form for a system of partial differential
equations on such fields.  The treatment would consist of a framework
sufficiently broad to encompass the systems found in
nature, but no broader. 
One would, for example, treat the initial-value formulation once
and for all within this broad framework, with the
formulations for individual physical systems emerging
as special cases.  In a similar way, one would treat---within
a quite general context---constraints, the
geometrical character of physical fields, how
some systems require other fields as a background,
how interactions operate, etc.  The goal of such a
treatment would be to get a better grip on the structural 
features of the partial differential equations of physics.
Here are two examples of issues on which such a treatment
might shed light.  What, if any, is the physical basis on
which the various fields on the manifold $M$ are grouped into
separate physical systems?  Thus, for instance, the fields
($n$, $F_{ab}$, $\rho$, $u^a$) are grouped into
($n$, $\rho$, $u^a$) (a perfect fluid), and ($F_{ab}$)
(electromagnetic field).  By ``physical basis", we mean
in terms of the dynamical equations on these fields.
A second issue is this:  How does it come about that
the fields of general relativity are singled out as
those for which diffeomorphisms on $M$ are gauge?  On
its face, this singling out seems surprising, for the diffeomorphisms
act equally well on all the physical fields on $M$.

We shall here discuss, in a general, systematic way, the
structure of the partial differential equations describing
physical systems.  We take it as given that there is a
fixed, four-dimensional manifold $M$ of ``space-time events",
on which all the action takes place.  Thus, for instance, we are not
considering discrete models.  Further, physical systems
are to be described by certain ``fields" on M.  These may be 
more general than mere tensor fields:  Our framework will
admit spinors, derivative operators, and perhaps other
field-types not yet thought of.  But we shall insist---largely
for mathematical convenience---that the set of field-values
at each space-time point be finite-dimensional.  We shall
further assume that these physical fields are subject to systems
of partial differential equations.  That is, we assume, 
among other things, that ``physics is local in $M$".  Finally,
we shall demand that these partial differential equations
be first-order (i.e., involving no higher than
first space-time derivatives of the fields), and quasilinear
(i.e., linear in those first derivatives).  That the equations
be first-order is no real assumption:  Higher-order
equations can---and will---be cast into first-order
form by introducing new auxiliary fields.  (Thus,
to treat the Klein-Gordon equation on scalar field $\psi$,
we introduce an auxiliary vector field playing the role of
$\nabla\psi$.)  It is my sense that this is more than
a mere mathematical device:  The auxiliary
fields tend to have direct physical significance.  It is
not so clear what we are actually assuming when we demand
quasilinearity.  It is certainly possible to write down a
first-order system of partial differential equations 
that is not even close to being quasilinear
(e.g., $(\partial\psi/\partial x)^2 + (\partial\psi/\partial t)^2
=\psi^2$). 
But all known physical systems seem to be described by quasilinear
equations, and it is anyway hard to proceed without
this demand.  In any case, ``first-order, 
quasilinear" allows us to cast all the partial differential
equations into a convenient common form, and it is on this common
form that the program is based.  A case could be made that, at least on
a fundamental level, all the ``partial differential equations 
of physics" are hyperbolic---that, e.g., elliptic and parabolic
systems arise in all cases as mere approximations of hyperbolic systems.
Thus, Poisson's equation for the electric potential is
just a facet of a hyperbolic system, Maxwell's equations.

In Sect. 2, we introduce our general framework for systems of 
first-order, quasilinear partial differential equations for the
description of physical systems.  The physical fields become 
cross-sections of an appropriate fibre bundle, and it is on these
cross-sections that the differential equations are written.
So, for instance, the coefficients in these equations
become certain tensor fields on the bundle space.  This
framework, while broad in its reach, is not particularly useful for
explicit calculations.  The remaining sections describe
various structural features of these system of
partial differential equations.  A ``hyperbolization" (Sect. 3)
is a casting of the system of equations (or, commonly, a 
subsystem of that system) into what is called symmetric,
hyperbolic form.  To such a form there is applicable
a general theorem on existence and uniqueness of solutions.
This is the initial-value formulation.  The constraints
(Sect. 4) represent a certain subsystem of the full system,  
the equations of which  
play a dual role:  providing conditions
that must be satisfied by initial data, and leading to
differential identities on the equations themselves.  The
constraints are integrable if these ``differential identities"
really are identities; and complete if the constraint
subsystem, together with the subsystem involved in the
hyperbolization, exhausts the full system of equations. 
 The geometrical character of the physical
fields has to do with how they ``transform", i.e., with 
lifting diffeomorphisms on $M$ to diffeomorphisms on the bundle
space (Sect. 5).  Combining all the systems of
physics into one master bundle $B$, then the full set
of equations on this bundle will be $M$-diffeomorphism invariant.
  This diffeomorphism invariance requires 
an appropriate adjustment in the initial-value formulation 
for this combined system. 
Finally, we turn (Sect. 6) to the relationships between
the various physical fields, as reflected in their
differential equations. 
Physical fields on space-time can interact on two
broad levels:  dynamically (through their derivative-terms),
and kinematically (through terms algebraic in the fields).
 Roughly speaking, two fields are part of
the same physical system if their derivative-terms
cannot be separated into individual equations; and one field is a background for
another if the former appears algebraically in the
derivative-terms of the latter. The kinematical (algebraic) interactions are the
more familiar couplings between physical systems.

It is the examples that give life to this general theory.
We have assembled, in Appendix A, a variety of
standard examples of physical systems:  the fields, the equations,
the hyperbolizations, the constraints, the background fields,
the interactions, etc.  We shall refer to this material
frequently as we go along.  Thus, this is not the standard
type of appendix (to be read later, if at all, by those
interested in technical details), but rather is an integral
part of the general theory.  Indeed, it might be well
to review this material first as a kind of introduction.  Appendix B
contains a statement and an outline of the proof of the theorem
on existence and uniqueness of solutions of symmetric,
hyperbolic systems of partial differential equations.
 
All in all, this subject forms a pleasant
comingling of analysis, geometry, and physics.
 
\section{Preliminaries}

Fix, once and for all, a smooth, four-dimensional manifold\footnote{We
take $M$ to be connected, paracompact, and Hausdorff.} $M$.  The
points of $M$ will be interpreted as the events of space-time, and,
thus, $M$ itself will be interpreted as the space-time manifold.
We do not, as yet, have a metric, or any other geometrical structure,
on $M$.

We next wish to introduce various types of ``fields" on $M$.  To this
end, let $b\stackrel{\pi}{\rightarrow}M$ be a smooth fibre bundle\footnote{See, e.g., Steenrod, {\em The Topology of Fibre Bundles} (Princeton
University Press, Princeton, l954). Note that, in contrast to what is
done in this reference, we introduce no Lie group acting on $b$.} over $M$.  That
is, $b$ is some smooth manifold (called the {\em bundle manifold})
and $\pi$ is some smooth mapping (called the {\em projection
mapping}); and these are such that locally (in $M$) $b$ can be written
as a product in such a way that $\pi$ is the projection onto one 
factor\footnote{This means, in more detail, that, given any point $x\in
M$, there exists an open neighborhood $U$ of $x$, a manifold $F$,
and a diffeomorphism $\zeta$ from $U\times F$ to $\pi^{-1}[U]$ such that
$\pi\circ\zeta$ is the projection of $U\times F$ to its
first factor.}.  An example is the tangent bundle of $M$:  Here, $b$ is the
eight-dimensional manifold of all tangent vectors at all points of $M$,
and $\pi$ is the mapping that extracts, from a tangent vector at a point
of $M$, the point of $M$.  That the local-product condition holds, in
this example, 
is seen by expressing tangent vectors in terms of their components
with respect to a local basis in $M$.  Returning to the general
case, for any point $x$ of $M$, the {\em fibre} over $x$ is the set of points ${\pi^{-1}}(x)$, i.e., the set of points
$\kappa \in b$ such that $\pi ( \kappa ) = x$.  It follows from the
local-product condition that each fibre is a smooth submanifold
of $b$, and that all the fibres are diffeomorphic with each
other.  In the example of the tangent bundle, for instance, the
fibre over point $x \in M$ is the set of all tangent vectors at
$x$.  Next, let $A$ be any smooth submanifold of $M$. 
A {\em cross-section}
 over $A$ is a smooth mapping $A\stackrel{\phi}{\rightarrow} b$ 
such that $\pi \circ \phi$ is the identity mapping on $A$.
Thus, a cross-section assigns, to each point $x$ of $A$, a point
of the fibre over $x$.  Typically, $A$ will be of dimension four
(i.e., an open subset of $M$), or three.
 
We interpret the fibre over $x$ as the space of allowed physical
states at the space-time point $x$, i.e., as the space of
possible field-values at $x$.  Then the bundle manifold
$b$ is interpreted as the space of all field-values at all
points of $M$.  A cross-section over submanifold $A$ becomes a field, defined at
the points of $A$.  In most, but not all, 
examples (Appendix A) $b$ will be a tensor bundle.  Thus, for
electromagnetism $b$ is the ten-dimensional manifold of all
antisymmetric, second-rank tensors at all points of $M$.  For
general relativity, by contrast, $b$ is the
fifty-four-dimensional 
 manifold a point of which is comprised of a point of $M$, a 
Lorentz-signature metric at that point, and a torsion-free derivative
operator at that point.  In both of these examples, the 
projection $\pi$ merely extracts the point of $M$. 

It is convenient to introduce the following notation.  Denote
tensors in $M$ by lower-case Latin indices; and tensors in
$b$ by lower-case Greek indices.  Then, at any point
$\kappa \in b$, we may introduce mixed tensors, where Latin
indices indicate tensor character in $M$ at $\pi (\kappa )$,
and Greek indices tensor character in $b$ at $\kappa$.  For
example, the derivative of the projection map is written
${(\nabla\pi)_{\alpha}}^a$, i.e., it sends tangent vectors
in $b$ at point $\kappa$ to tangent vectors in $M$ at
$\pi ( \kappa )$.  The derivative of a cross-section,
$\phi$, over a four-dimensional region of $M$ is written
${(\nabla \phi )_a}^{\alpha}$; and we have, from the
defining property of a cross-section,  
\begin{equation}
(\nabla\phi)_a{^{\alpha}}(\nabla\pi)_{\alpha}{^b}
=\delta_a{^b}.
\label{2.0}\end{equation}
 
A vector $\lambda ^ {\alpha}$ at $\kappa \in b$ is called 
{\em vertical} if it is tangent to the fibre through
$\kappa$, i.e., if it satisfies $\lambda ^{\alpha}{ (\nabla
\pi )_{\alpha}}^{b} = 0$.  Elements of the space of vertical
vectors at $\kappa$ will be denoted by primed Greek 
superscripts.  Thus, $\lambda^{\alpha'}$ means ``$\lambda$ is
a tangent vector in $b$, a vector which, by the way, is vertical".  Elements of
the space dual to that of the vertical vectors will be denoted
by primed Greek subscripts.  Thus, $\mu _{\alpha'}$ means
``$\mu$ is a linear mapping from vertical vectors in $b$ 
 to the reals".  More generally, these primed
indices may appear in mixed tensors.  Note that we may freely
remove primes from superscripts (i.e., ignore the
verticality of an index), and add primes to subscripts
(i.e., restrict the mapping from all tangent vectors
to just vertical ones), but not the reverses.  As an
example of this notation, we have:  ${(\nabla \pi )_{\alpha '}}
^{a} = 0$.

To illustrate these ideas, consider electromagnetism.  Then a typical
point of the bundle manifold $b$ is $\kappa=(x,F_{ab})$, where $x$
is a point of $M$ and $F_{ab}$ is an antisymmetric tensor at $x$.
A tangent vector $\lambda^{\alpha}$ in $b$ at $\kappa$ can be 
represented as
an ``infinitesimal change\footnote{By ``infinitesimal change in the point
of...", we mean ``tangent vector to a curve in...".} in both the point $x$ of $M$ and the
antisymmetric tensor $F_{ab}$".  Given such a $\lambda^{\alpha}$,
the combination $\lambda^{\alpha}{(\nabla\pi)_{\alpha}}^a$ is that
tangent vector in $M$ at $x$ represented by just the ``change
in $x$"-part of $\lambda^{\alpha}$ (ignoring the
``change in $F_{ab}$"-part). 
 Such a $\lambda^{\alpha}$
is vertical provided its ``change in x" 
vanishes---so, a vertical vector is represented simply as an infinitesimal change
in the antisymmetric tensor $F_{ab}$, with $x$ fixed.  In this example,
we might introduce the field $\mu_{\alpha'ab} = \mu_{\alpha'[ab]}$ 
on $b$,  
which takes any such vertical vector, $\lambda^{\alpha'}$, and 
returns, as $\lambda^{\alpha'}\mu_{\alpha'ab}$, the change in
the tensor $F_{ab}$ at $x$.

A {\em connection} on fibre bundle $b\stackrel{\pi}{\rightarrow}M$
is a smooth field ${\gamma_a}^{\alpha}$ on $b$ satisfying
${\gamma_a}^{\alpha}{(\nabla\pi)_\alpha}^b = {\delta_a}^b$.  Given
a connection ${\gamma_a}^{\alpha}$, those vectors at
$\kappa{\in}b$ that can be written in the form
$\xi^a{\gamma_a}^{\alpha}$ for some $\xi^a$ are 
called {\em horizontal}.  Of course, there exist many possible
connections, and so many such notions of ``horizontality".
It follows directly from these definitions that, fixing
a connection, every tangent vector in $b$ at $\kappa$
can be written, uniquely, as the sum of a horizontal and a 
vertical vector, i.e., that every vector can be split into 
its horizontal and vertical parts.  We may incorporate this
observation into the notation by allowing ourselves the
operations with primes that were previously prohibited:
In the presence of a fixed connection, $\gamma_a{^\alpha}$, we may affix a prime
to a Greek superscript (by taking the vertical projection);
and, in a similar way, we may remove a prime from a
Greek subscript.  For example, we have 
$\gamma_a{^{\alpha'}}=0$.  Note that in every case the removal and
subsequent affixing of a prime leaves a tensor unchanged
(but not so for affixing a prime and its subsequent removal.)

Again consider, as an example, the case of electromagnetism.
(Any other (nonscalar) tensor bundle would be similar.) 
Fix any smooth derivative operator $\nabla_a$ on the
manifold $M$.  Then this $\nabla_a$ gives rise to a
connection $\gamma_a{^{\alpha}}$ on $b$, in
the following manner. 
For $\kappa = (x,F_{ab})$ any point of $b$, and 
$\xi^a$ any tangent vector in $M$ at $x$, let
$\lambda^{\alpha}= \xi^a\gamma_a{^{\alpha}}$ be that
tangent vector in $b$ at $\kappa$ represented
 as follows:
``The infinitesimal change in $x$ is that dictated
by $\xi^a$, while the infinitesimal change in $F_{ab}$
is that resulting from parallel transport, via $\nabla_a$,
of $F_{ab}$ from $x$ along $\xi^a$."  We thus specify the
combination $\xi^a\gamma_a{^{\alpha}}$ for every $\xi^a$,
and so the tensor $\gamma_a{^{\alpha}}$ itself.  Note that
 we have
$\lambda^{\alpha}{(\nabla\pi)_{\alpha}}^a = \xi^a$,
which shows that the ${\gamma_a}^{\alpha}$ so defined is 
indeed a connection.  So, the horizontal vectors
at $\kappa$ in this example are those for which ``the infiniestimal
change in $F_{ab}$ is exactly that resulting from parallel
transport".  Clearly, every tangent vector in $b$ can
be written, uniquely, as the sum of a vertical vector and such a horizontal
vector.  While every derivative
operator on $M$ gives rise, as above, to a connection
on $b$, there are many other connections on $b$
(corresponding roughly to ``non-linear parallel
transport").

We shall not routinely make use of a connection in what
follows, for two reasons.  First, for some fields, such as
the derivative operator of general relativity, we have no
natural connection.  Second, even when there is a natural
connection (e.g., for electromagnetism), 
 that connection
will itself be a dynamical variable.  It is awkward having 
one dynamical field  playing a crucial role in
the kinematics of another. 

We now wish to write down a certain class of partial
differential equations on cross-sections.  To this end,
let ${{k_A}^m}_{\alpha}$ and $j_A$ be smooth fields
on $b$.  Here, the index ``$A$" lives in some, as yet
unspecified, vector space.  Normally, this vector space will
be some tensor product involving tensors in $M$ and in
$b$, i.e., ``$A$" will merely stand for some combination
of Latin and Greek indices.  But, at least in principle,
this could be some newly constructed vector space attached 
to each point of $b$, in which case we would have to 
introduce a new fibre bundle, with base space $b$, to
house it.  Consider now the partial differential equation
\begin{equation}
{{k_A}^m}_{\alpha}{(\nabla\phi)_m}^{\alpha} + j_A = 0,
\label{2.1}
\end{equation}  
where $U\stackrel{\phi}{\rightarrow}b$ is a smooth
cross-section over some open subset $U$ of $M$.
This equation is to hold at every point $x{\in}U$, where
$k$ and $j$ are evaluated ``on the cross-section", i.e.,
at $\phi(x)$.  Note that this is a first-order
equation on the cross-section, linear in its first
derivative.  The ``number of unknowns" at each point is the
dimension of the fibre; the ``number of equations" the
dimension of whatever is the vector space in which the
index ``$A$" lives.  The coefficients in this
equation, ${{k_A}^m}_{\alpha}$ and $j_A$, are functions
on the bundle manifold $b$, i.e., these coefficients
may ``depend on both the point of $M$ and the
field-value $\phi$".

Apparently, every system of partial differential equations
describing a physical system in space-time can be cast into
the form of Eqn. (\ref{2.1}).  Various examples are given in
Appendix A.  Many, such as those for a perfect fluid,
the electromagnetic field, or the charged Dirac particle, are already
packaged in the appropriate form.  Others must be brought
into this form by introducing auxiliary fields.
In the Klein-Gordon case, for example, we must augment
the scalar field $\psi$ by its space-time 
derivative, $\psi_a$, resulting in a bundle
space with five-dimensional fibres.  We then obtain,
on $(\psi,\psi_a)$, a first-order system of
equations of the form (\ref{2.1}).  For general relativity,
the fibre over $x{\in}M$ consists of pairs 
$(g_{ab},\nabla_a)$, where $g_{ab}$ is a Lorentz-signature 
metric and $\nabla_a$ a torsion-free
derivative operator at $x$.  The curvature tensor
arises in (\ref{2.1}) as the derivative of the derivative operator.
Let us agree that {\em all} first-order equations on the
fields (even those that follow from differentiating other
equations) are to be included in (\ref{2.1}).  Thus, for
example, Eqn. (\ref{A.7}) is included for the
Klein-Gordon system. 
Note that the only structure we are imposing on the physical
fields at this stage is a differentiable structure, as carried by
the manifold $b$.  If you wish to utilize any additional features 
on these fields---e.g., the ability to add fields, to multiply them
by numbers, to multiply them by each other, etc.---then this 
must be introduced, separately and explicitly, as additional structure
on the bundle space $b$.  For example, for electromagnetism, but 
not for a perfect fluid, each fibre has the
additional structure of a vector space

The present formulation of partial differential equations
carries with it a certain gauge freedom.  Let
${{\lambda_A}^m}_b$ be any smooth field on $b$. 
Then Eqn. (\ref{2.1}) remains invariant under adding to
${{k_A}^m}_{\alpha}$ the expression
${{\lambda_A}^m}_b{(\nabla\pi)_{\alpha}}^b$, and
at the same time to $j_A$ the expression
$-{{\lambda_A}^m}_m$.  That is, the solutions
$\phi$ of (\ref{2.1}) before these changes in $k$ and
$j$ are precisely the same as the solutions after.
Note that ${{k_A}^m}_{\alpha'}$, (i.e., what results
from contracting ${{k_A}^m}_{\alpha}$ only with 
vertical $v^{\alpha}$) is gauge-invariant.  Furthermore,
this tensor exhausts the gauge-invariant information,
in the following sense:  Given any field 
$\hat{k}_A{^m}{_{\alpha}}$ satisfying
$\hat{k}_A{^m}{_{\alpha'}} = {{k_A}^m}_{\alpha'}$,
then there exists one and only one gauge transformation
that sends ${{k_A}^m}_{\alpha}$ to 
$\hat{k}_A{^m}{_{\alpha}}$.  This gauge
freedom reflects the idea that ``the horizontal part" 
of the $\alpha$-contraction in (\ref{2.1}) does not really
involve the derivative of the cross-section, by virtue
of the identity (\ref{2.0}). 
 Thus, the components of
$k_A{^m}{_{\alpha}}$ that participate in
this part of the $\alpha$-contraction
are not contributing to the dynamics.
It would be
most convenient if we could somehow circumvent this 
gauge freedom, e.g., by rewriting Eqn. (\ref{2.1}) to involve
only the gauge-invariant part, ${{k_A}^m}_{\alpha'}$,
of ${{k_A}^m}_{\alpha}$.  Unfortunately, this cannot
be done in any natural way in general.  But it can be
done in the presence of some fixed connection,
${\gamma_a}^{\alpha}$, on the bundle $b$.
In fact, given a connection, we may always achieve
through gauge a ${{k_A}^m}_{\alpha}$ that is ``vertical
in $\alpha$" in the sense that it annihilates every
horizontal vector $h^{\alpha}$.  Furthermore, this
requirement on ${{k_A}^m}_{\alpha}$ completely
exhausts the gauge freedom.  Indeed, given 
${{k_A}^m}_{\alpha}$ and connection $\gamma_a{^{\alpha}}$, then the gauge transformation with 
${{\lambda_A}^m}_b = - {\gamma_b}^{\alpha}{{k_A}^m}_{\alpha}$,
uniquely, does the job.  It will sometimes be convenient,
when a connection is available, to exploit this gauge-choice.

\section{Hyperbolizations}

A key feature of the partial differential equations of physics is
their initial-value formulation, i.e., their formulation in terms
of initial data and ``time"-evolution.  It turns out that this
formulation
can be carried out in a rather general setting.  This  is
the subject of the present, and much of the following, section.

Fix a partial differential equation of the form (\ref{2.1}), so
we have in particular fixed smooth fields $k_A{^m}{_\alpha}$
and $j_A$ on $b$.  By a {\em hyperbolization} of Eqn. (\ref{2.1}),
we mean a smooth field $h^A{_{\alpha'}}$ on $b$ such that i)
the field $h^A{_{\alpha'}}k_A{^m}{_{\beta'}}$ on $b$ is symmetric
in $\alpha',\beta'$; and ii) for each point $\kappa\in b$, there
exists a covector $n_m$ in $M$ at $\pi(\kappa)$ such that the
tensor $n_m h^A{_{\alpha'}}k_A{^m}{_{\beta'}}$ at $\kappa$ is
positive-definite.  Note that the definition involves only
$k_A{^m}{_{\alpha'}}$, and neither $j_A$ nor the rest of $k$.
Thus, in particular, the definition is gauge-invariant.  Note
also that the hyperbolizations at a point $\kappa\in b$ form an
open subset of a vector space.  For $h^A{_{\alpha'}}$ any
hyperbolization, and $v^{\alpha'}$ any nonzero (vertical)
vector at a point, the combination $v^{\alpha'}h^A{_{\alpha'}}$
at that point must be nonzero. (This follows, contracting the
positive-definite tensor in ii) with $v^{\alpha'}v^{\beta'}$.)
But this implies, in turn, that the dimension of the space of
 equations in (\ref{2.1}) (that of the index ``$A$") must be
greater than or equal to the dimension of the space of
unknowns (that of the index ``$\alpha'$").  So, if this dimensionality
criterion fails, then there can be no hyperbolization.  But 
suppose this criterion is satisfied:  Can we then guarantee
a hyperbolization?  The answer is no.  In fact, there is no
known, practical procedure, given a general partial
differential equation (\ref{2.1}), for finding its hyperbolizations,
or, indeed, for even determining whether or not one exists. 
(This is essentially a little algebra problem:  Given a tensor
$k_A{^m}{_{\alpha'}}$ at a point, what are the tensors
$h^A{_{\alpha'}}$ at that point with $h^A{_{\alpha'}}k_A{^m}{_{\beta'}}$
symmetric?)  In practice, hyperbolizations are found, 
in sufficiently low dimensions, by solving explicitly the algebraic equations
inherent in i) and ii) ; and, in higher
dimensions, by guessing.  Physical considerations
frequently suggest candidates.

Consider again the example of electromagnetism (Appendix A).  We
have already remarked that, at point $\kappa= (x,F_{ab})$ of
the bundle space $b$, a typical vertical vector, which we
now write $\delta\phi^{\alpha'}$, is represented by an
infiniesimal change, $\delta F_{ab}$, in the electromagnetic
field at $x$.  Since the left sides of Maxwell's equations,
(\ref{A.1}) and (\ref{A.2}), consist of a vector and a third-rank,
antisymmetric tensor, the index ``$A$" lies in the eight-dimensional
vector space of such objects.  That is, a typical vector
in this space can be written $\sigma^A= (s^a, s^{abc})$, with
$s^{abc}=s^{[abc]}$.  (Note that, since dim ``$A$" = $8 > 6$
= dim ``$\alpha'$", our dimensionality criterion above is 
satisfied.)  The fields $k_A{^m}{_\alpha}$ and $j_A$ are to
be read off by comparing Maxwell's equations, (\ref{A.1}) and
(\ref{A.2}), with the general partial differential equation
(\ref{2.1}).  We thus obtain
\begin{equation}
k_A{^m}{_{\beta'}}\sigma{^A}n_m\delta \hat{\phi}{^{\beta'}}
=s^a(n^b\delta \hat{F}_{ab})+s^{abc}(n_{[a}
\delta\hat{F}_{bc]}).
\label{3.1}
\end{equation}
Here, we have represented $k_A{^m}{_{\beta'}}$ by giving the
scalar that results from contracting away its indices,
on vectors $\sigma^A$, $n_m$, and $\delta \hat{\phi}^{\beta'}$.
The field $j_A$ of (\ref{2.1}), on the other hand, depends on gauge.
If we choose for the gauge that determined by the derivative
operator $\nabla_a$ on $M$ used in Maxwell's
equations (\ref{A.1}), (\ref{A.2}), then we
have $j_A=0$.  Now fix any vector $t^a$ at $x$, and consider
the tensor $h^A{_{\alpha'}}$ given, in Eqn. (\ref{A.3}), as 
the $A$-index vector that results from the
contraction $h^A{_{\alpha'}}\delta\phi^{\alpha'}$.  
Substituting this vector for $\sigma^A$ in (\ref{3.1}), we obtain
\begin{eqnarray}
h^A{_{\alpha'}}\delta\phi^{\alpha'}k_A{^m}{_{\beta'}}
n_m\delta \hat{\phi}^{\beta'}&=&
\delta F^a{_m}t^m(n^b\delta \hat{F}_{ab})-\frac{3}{2}t^{[a}
\delta F^{ab]}(n_{[a}\delta \hat{F}_{bc]})\nonumber\\
&=& 2t^an^b[\delta F_{(a}{^m}\delta \hat{F}_{b)m}
- 1/4 g_{ab} \delta F_{mn} \delta \hat{F}^{mn}]. 
\label{3.2}
\end{eqnarray}
It now follows, provided only that the vector
$t^a$ is chosen timelike, that the $h^A{_{\alpha'}}$ of
(\ref{A.3}) is a hyperbolization.  Indeed, condition
i) follows from the fact that the last expression
in (\ref{3.2}) is symmetric under interchange of
$\delta F_{ab}$ and $\delta \hat{F}_{ab}$; and
condition ii) follows from the fact that, whenever
$n_m$ is timelike with $t^m n_m < 0$, the last
expression in (\ref{3.2}) defines a positive-definite
quadratic form in $\delta F_{ab}$.  Thus, every
timelike vector field $t^a$ on $M$ gives rise to a
hyperbolization of Maxwell's equations.  In fact,
this family exhausts the hyperbolizations in the
Maxwell case.

The situation is similar for many other physical
examples. (See Appendix A.)  Thus, the hyperbolizations of
the Klein-Gordon equation are characterized by two vector
fields on $M$; and those for the perfect-fluid equation by
two scalar fields.  Even dissipative fluids\footnote{See, e.g., 
I.S. Liu, I. Muller, T. Ruggeri, {\em Ann Phys (NY) 169}, 191
(1986); R. Geroch, L. Lindblom, {\em Ann Phys (NY) 207}, 394 (1991);
R. Geroch, {\em J. Math. Phys. 36}, 4226 (1995).} can
be described by equations admitting a hyperbolization.
There are only two physical examples,
as far as I am aware, for which there exist no
hyperbolization.  One is Einstein's equation, for which
the lack of a hyperbolization is related to the 
diffeomorphism-invariance of the theory; and the other is
dust.  We shall return to each of these examples later.

Fix a hyperbolization, $h^A{_{\alpha'}}$, of Eqn. (\ref{2.1}).
For each point $\kappa\in b$, denote by $s_{\kappa}$ the
collection of all covectors $n_m$ in $M$ at
$\pi(\kappa)$ such that the tensor $n_m h^A{_{\alpha'}}
k_A{^m}{_{\beta'}}$ is positive-definite.  Then $s_{\kappa}$
is a nonempty (by condition ii)), open, convex cone.
The physical interpretation of these cones will turn out
to be the following.  The tangent vectors $p^a$ in $M$
at $\pi(\kappa)$ such that $p^an_a > 0$ for all 
$n_a\in s_{\kappa}$ represent the ``signal-propagation
directions" of the physical field. Note that these 
$p^a$ form a closed, nonempty, convex cone at each point,
the ``dual cone" of $s_{\kappa}$.  These cones depend not
only on the point $x$ of $M$, but also in general
on the value of the field at $x$, i.e., on where we are
in the fibre over $x$.  In cases in which there is more 
than one hyperbolization, these cones could also depend on
which hyperbolization has been selected.  But it turns
out that, for most physical examples, these cones are essentially
independent of hyperbolization.  
Thus, 
in the case of electromagnetism, the signal propagation
directions $p^a$ consist of all
timelike and null vectors lying in one of the two
halves of the light cone.  In the case of a perfect
fluid, the $p^a$ form the ``sound
cone".  Is it possible to isolate, via a definition, the
crucial algebraic feature of $k_A{^m}{_{\alpha'}}$
in such physical examples that is responsible for
hyperbolization-independent cones? 

Suppose that, included among the various physical
fields on $M$ is a spacetime metric, $g_{ab}$.  In that
case, we say that the system (\ref{2.1}) is
{\em causal} if all the signal-propagation directions
are timelike or null.  This is equivalent to the condition
that each $s_{\kappa}$ includes all timelike vectors
lying within one of the two halves of the light
cone.  A perfect fluid, for example, is causal
provided its sound speed, $\frac{\partial p}{\partial\rho}
+ \frac{n}{\rho+p}\frac{\partial p}{\partial n}$, 
 does not
exceed the speed of light. 

Fix a hyperbolization $h^A{_{\alpha'}}$ of Eqn.
(\ref{2.1}).  This hyperbolization leads, as we now
explain, to an initial-value formulation.  By
{\em initial data} we mean a smooth, three-dimensional
submanifold $S$ of $M$, together with a smooth cross-section,
$S\stackrel{\phi_0}{\rightarrow}b$, over $S$, 
such that, for every point $x\in S$, a normal $n_m$
to $S$ at $x$ lies in the cone $s_{\phi_0(x)}$.  In
other words, we must specify the physical state of
the system at each point of the three-dimensional
manifold $S$, in such a way that, at every point of $S$,
all signal-propagation directions are transverse
to $S$.  Note that the role of the cross-section,
$\phi_0$, in this definition is to determine the
cone within which the normal to $S$ must lie.  Thus,
a change of cross-section, keeping $S$ fixed, could
destroy the initial-data character of $(S,\phi_0)$. 
(Changing the hyperbolization could, in principle, also
change the initial-data character, but, as we remarked earlier,
it generally does not.) As an example of these definitions 
we have:  If we have on $M$
a spacetime metric $g_{ab}$ with respect to which
(\ref{2.1}) is causal, then every $(S,\phi_0)$, with
$S$ spacelike, is initial data.

  We may now summarize
the fundamental existence-uniqueness theorem as
follows.  Given initial data $(S,\phi_0)$, there exists,
in a suitable neighborhood $U$ of $S$, one and only one
cross-section, $U\stackrel{\phi}{\rightarrow}b$, such
that i) $\phi | {_S} =\phi_0$ and ii)
\begin{equation}
h^A{_{\beta'}}[k_A{^m}{_\alpha}(\nabla\phi)_m{^\alpha}
+j_A]=0. 
\label{3.3}
\end{equation}
Condition i) ensures that the field $\phi$, specified over
the neighborhood $U$ of $S$, agree, on $S$ itself, with
the given initial conditions, $\phi_0$.  Condition ii)
ensures that the field $\phi$ satisfy a certain partial
differential equation derived from (\ref{2.1}) (specifically,
by contracting it with $h^A{_{\beta'}}$).  In short, the
theorem states that we can ``solve" the partial
differential equation (\ref{3.3}), uniquely, subject to any
given initial conditions.  There is given in Appendix B
a more detailed version of this theorem (including 
more information regarding the neighborhood
$U$), and a sketch of the proof.  This
version, in particular, supports our interpretation
of the cones $s_{\kappa}$ in terms of signal-propagation.

Since every solution of Eqn. (\ref{2.1}) is automatically
a solution of (\ref{3.3}), the theorem above guarantees
local uniqueness of the solutions of any system,
(\ref{2.1}), of partial differential equations admitting
a hyperbolization.  Thus, for most systems of interest
in physics, initial data lead to a unique local
solution.  Furthermore, if the hyperbolization
$h^A{_{\alpha'}}$ is invertible (which holds, by the
way, if and only if dim ``$A$" = dim ``$\alpha'$"), then
Eqn. (\ref{3.3}) is equivalent to Eqn. (\ref{2.1}).
In this case, e.g., for a perfect fluid, the theorem
also guarantees local existence of solutions of 
(\ref{2.1}).  But in many physical examples---electromagnetism
included---$h^A{_{\alpha'}}$ is not invertible---so part
of Eqn. (\ref{2.1}) is lost in the passage to
(\ref{3.3}). In these cases, we cannot guarantee, directly from the theorem,
local existence of solutions of Eqn. (\ref{2.1}).
The fate of these ``lost equations" is the subject
of the following section.

Let us now return briefly to the example of dust.
(See Appendix A.)  With the traditional choice of 
fields---$\rho$ (mass density) and $u^a$ (unit, 
timelike four-velocity)---the dust equations, 
(\ref{A.41}) and (\ref{A.42}), admit no hyperbolization.
This is perhaps surprising, for this system ``obviously" has
an initial-value formulation.  It turns out that, if we
introduce the auxiliary field $w_a{^b}=\nabla_a u^b$,
then the corresponding system of equations on this new set of 
fields, $(\rho,u^a,w_a{^b})$, does admit a
hyperbolization. It is not clear what, if any, is the
physical meaning of this modification.
Furthermore, the hyperbolization it produces is
apparently lost on coupling the dust with gravitation
via Einstein's equation.  (This behavior is a consequence
 of the appearance
of a Riemann tensor in the equations on
$(\rho, u^a, w_a{^b})$.)  
What is going on physically in this example? 

\section{Constraints}

Fix a partial differential equation of the form (\ref{2.1}),
so we have in particular fixed smooth fields $k_A{^m}{_\alpha}$
and $j_A$ on $b$.  While much of the material of this 
section finds application to the initial-value formulation,
we require at this stage no specific hyperbolization---nor, indeed,
even the existence of one.

A {\em constraint} at point $\kappa\in b$ is a tensor
$c^{An}$ at $\kappa$ such that
\begin{equation}
c^{A(n}k_A{^{m)}}{_{\alpha'}} = 0. 
\label{4.1}
\end{equation}
Note that the definition is gauge-invariant, and that the
constraints at $\kappa$ form a vector space.  A number of
examples, for various physical systems, is given in
Appendix A.  For instance, the equations for a perfect
fluid admit only the zero constraint; those for 
Klein-Gordon, a ten-dimensional vector space of
constraints; and those for general relativity an
eighty-four dimensional vector space.  Maxwell's equations, on the 
other hand, admit a two-dimensional vector space of 
constraints:  The general $c^{An}$ is given, in this case, by Eqn.
(\ref{A.4}), where $x$ and $y$ are arbitrary numbers.
To check that this $c^{An}$ does indeed satisfy (\ref{4.1}),
combine it with the
$k_A{^m}{_{\alpha'}}$ for Maxwell's equations
given by (\ref{3.1}), to obtain
\begin{equation}
c^{An}k_A{^m}{_{\alpha'}}\delta\hat{\phi}^{\alpha'}
= x\delta\hat{ F}^{nm} + y\epsilon^{nmbc}\delta \hat{F}_{bc}. 
\label{4.2}
\end{equation}
Now symmetrize both sides over $n,m$. 
  Each constraint, as we
shall see, plays two distinct roles:  It signals a 
differential condition that must be imposed on initial
data for Eqn. (\ref{2.1}), as well as a differential
identity involving Eqn. (\ref{2.1}).  In the case of
Maxwell's equations, for example, the first role is reflected
in the familiar ``spatial constraint equations",
$\bf{\nabla \cdot E = 0}$, $\bf{\nabla\cdot B = 0}$.
The second role is reflected in the fact that identities result
from taking 
the divergence and curl, respectively, of Maxwell's
equations, (\ref{A.1}) and (\ref{A.2}).

We begin with the first role.  Fix constraint $c^{An}$.
Let $U\stackrel{\phi}{\rightarrow}b$ be any solution
of Eqn. (\ref{2.1}), defined in open $U\subset M$, and let
$S$ be any three-dimensional submanifold of $U$.
Consider now the equation
\begin{equation}
n_ac^{Aa}[k_A{^m}{_{\alpha}}(\nabla\phi)_m{^{\alpha}}
+ j_A ] = 0,
\label{4.3}
\end{equation}
at points $x$ of $S$, where $n_a$ is a normal to $S$
at $x$ and the coefficients are evaluated at
$\kappa=\phi(x)$.  We first note that Eqn. (\ref{4.3})
holds on $S$, for it is a consequence of (\ref{2.1}).
We next claim that the left side of Eqn. (\ref{4.3})
involves only $\phi_0=\phi | {_S}$, i.e., only
$\phi$ restricted to $S$.  To see this, first note
that $\phi_0$ alone determines $(\nabla\phi)_m{^{\alpha}}$
at points of $S$ up to addition of a term of the 
form $n_mv^{\alpha'}$.  But such a term
annihilates $n_ac^{Aa}k_A{^m}{_{\alpha}}$, by the
defining equation, (\ref{4.1}), for a constraint,
and so does not contribute to the left side of
Eqn. (\ref{4.3}).  What we have shown, then,
is that Eqn. (\ref{4.3}) is a ``constraint
equation":  It is a differential equation on
cross-sections over $S$ that must be satisfied by every
restriction to $S$ of a solution of Eqn. (\ref{2.1}). 
In the Maxwell case, for example, the two
independent constraints give rise, via (\ref{4.3}),
to the vanishing of the divergence of the electric
and magnetic fields.  Note that $(S,\phi_0)$ above need 
not be initial data:  We have as yet introduced no
hyperbolization.

We next introduce a notion of ``sufficiently many" constraints.
We say the constraints are {\em complete} if, for any
point $\kappa\in b$ and any nonzero covector $n_n$ at
$\pi(\kappa)\in M$, we have
\begin{equation}
{\rm dim}(c^{An}n_n) +{\rm dim}(v^{\alpha'})={\rm dim} (\sigma^A).
\label{4.4}
\end{equation}
The first term\footnote{Note that this term can be---and in
examples (such as Klein-Gordon) frequently is---less than the dimension
of the vector space of constraints.} is the dimension of the space of all 
vectors of the indicated form, as $c^{An}$ runs over
all constraints at $\kappa$.  The second term is the
dimension of the space of vertical vectors, i.e.,
the dimension of the fibres.  The last term is the dimension
of the space of equations represented by (\ref{2.1}).
Eqn. (\ref{4.4}) means, roughly speaking, that there
are at least as many equations as unknowns in
Eqn. (\ref{2.1}), and that any excess is taken
up entirely by constraint equations, (\ref{4.3}).
This interpretation will be made more precise 
shortly.

The constraints are complete for the vast majority of 
physical examples.  (See Appendix A.)  Thus, Eqn.
(\ref{4.4}) reads, for the perfect-fluid equations,
``$0 + 5 = 5$"; for Maxwell's equations, ``$2 + 6 = 8$";
for the Klein-Gordon equations, ``$6 + 5 = 11$".  For Einstein's
equation, the constraints are not complete:  Eqn. (\ref{4.4})
reads ``$64+50 = 110$".  This, as we shall see later, is
related to the diffeomorphism-invariance of
the theory\footnote{We remark that there exist examples (though
apparently no physically interesting ones) of a tensor
$k_A{^m}{_\alpha'}$ admitting a hyperbolization, but
whose constraints are not complete.}. 
  Is there
some simple characterization of those tensors
$k_A{^m}{_{\alpha'}}$ that yield complete constraints?

The second role of a constraint is in signaling a differential
identity involving Eqn. (\ref{2.1}).  The idea here is very
simple.  Eqn. (\ref{2.1}) is to hold at every point $x$
of some open subset $U$ of $M$.  Taking the $x$-derivative,
$\nabla_n$, of this equation, and contracting with any
constraint $c^{An}$, we obtain an equation involving the first
and second derivatives of the cross-section.  But, as it turns out, the
second-derivative term drops out, by virtue of (\ref{4.1}),
and so we are left with an algebraic---in fact, quadratic---equation
in the first derivative, $(\nabla\phi)_m{^{\alpha}}$,
of the cross-section.  That is, we obtain an integrability
condition for Eqn. (\ref{2.1}).  If this integrability condition
holds as an algebraic consequence of Eqn. (\ref{2.1}), then we
say our constraint is integrable.

Unfortunately, all this becomes somewhat more complicated when
written out explicitly.  Fix any (torsion-free) derivative
operator, $\nabla_{\alpha}$, on the manifold $b$, such that
the derivative of every vertical vector field is vertical.  (Such
always exists, at least locally, by the local-product character
of the fibre bundle.)  Extend\footnote{This is done as follows.
Any field $\xi^m$ can be written in the form
$(\nabla\pi)_{\beta}{^m}\xi^{\beta}$, uniquely but for the
freedom to add to $\xi^{\beta}$ any vertical vector field.  Now
define
$\nabla_{\alpha}\xi^m=(\nabla\pi)_{\beta}{^m}\nabla_{\alpha}\xi^{\beta}$,
noting that the right side is invariant under this freedom.  Note
that this extension of $\nabla_{\alpha}$ to fields with Latin
indices is unique.} this operator to mixed fields on $b$ by 
demanding $\nabla_{\beta}(\nabla\pi)_{\alpha}{^m}=0$.  Then
the operator ``derivative along the cross-section" is 
$(\nabla\phi)_n{^{\alpha}}\nabla_{\alpha}$.  Applying this
operator to (\ref{2.1}), and contracting with any constraint
$c^{An}$, we obtain
\begin{equation}
c^{An}(\nabla_{\beta}k_A{^m}{_{\alpha}})(\nabla\phi)_n{^{\beta}}
(\nabla\phi)_n{^{\alpha}}+c^{An}(\nabla_{\beta}j_A)
(\nabla\phi)_n{^{\beta}} = 0.
\label{4.5}\end{equation}
In the derivation of Eqn. (\ref{4.5}), there arises initially
the term $[c^{An}k_A{^m}{_{\alpha}}]$ $[(\nabla\phi)_n{^{\beta}}
\nabla_{\beta}(\nabla\phi)_m{^{\alpha}}]$, involving the
second derivative of the cross-section.  To see that this term
vanishes, first note that the index ``$\alpha$" in the
second factor is vertical (contracting with $(\nabla\pi)_{\alpha}
{^s}$), and so only the antisymmetrization of this factor
over $n,m$ contributes (by definition of a constraint),
yielding zero (by the torsion-free character of $\nabla_{\alpha}$).
Eqn. (\ref{4.5}) is our integrability condition.  We say
that constraint $c^{An}$ is {\em integrable} if Eqn.
(\ref{4.5}) is an algebraic consequence of Eqn. (\ref{2.1}).
What this means, in more detail, is that the left side
of Eqn. (\ref{4.5}) is some multiple 
of the left side of
(\ref{2.1}) plus some multiple
of the difference
between the two sides of the identity (\ref{2.0}), where
each of these two multiplying factors is an expression linear in $(\nabla\phi)_a
{^{\alpha}}$.   Writing
this out and equating powers of $(\nabla\phi)_a{^{\alpha}}$,
 we conclude:  Constraint
$c^{An}$ is integrable if and only if there exist tensors
$\sigma^{Am}{_{\alpha}}$ and $\sigma^a{_b}{^m}{_{\alpha}}$, 
with $\sigma^a{_a}{^m}{_{\alpha}}=0$, such that
\begin{eqnarray} 
&&\nabla_{\alpha}(c^{Am}\tilde{k}_A{^n}{_{\beta}})+
\nabla_{\beta }(c^{An}\tilde{k}_A{^m}{_{\alpha}})\nonumber\\
&&\ \ =\sigma^{Am}{_{\alpha}}\tilde{k}_A{^n}{_{\beta}}+
\sigma^{An}{_{\beta}}\tilde{k}_A{^m}{_{\alpha}}
+\sigma^m{_s}{^n}{_{\beta}}(\nabla\pi)_{\alpha}{^s}
+\sigma^n{_s}{^m}{_{\alpha}}(\nabla\pi)_{\beta}{^s},
\label{4.6}\end{eqnarray}
where we have set $\tilde{k}_A{^m}{_{\alpha}} =
k_A{^m}{_{\alpha}} + \frac{1}{4} j_A(\nabla
\pi)_{\alpha}{^m}$.  Applying a prime to both
``$\alpha$" and ``$\beta$" in this equation, and using
(\ref{4.1}), we obtain
\begin{equation}
2\nabla_{[\alpha'}(c^{Am}k_{|A|}{^n}{_{\beta']}}) =
\sigma^{Am}{_{\alpha'}}k_A{^n}{_{\beta'}}+
\sigma^{An}{_{\beta'}}k_A{^m}{_{\alpha'}}.
\label{4.7}\end{equation} 
This part of (\ref{4.6}) is manifestly gauge-invariant
(involving only $k_A{^m}{_{\alpha'}}$, and not $j_A$
or the rest of $k$), and independent of the derivative
operator $\nabla_{\alpha}$ (involving only the
``vertical curl").  What remains of Eqn. (\ref{4.6}) is essentially
one scalar relation, expressing the divergence of
$c^{An}j_A$ in terms of other fields.  Is there some
simple way to write this remaining relation, e.g., a way
that separates its physical content from the gauge freedom
inherent in ($k$, $j$), $\nabla_{\alpha}$, and the $\sigma$'s?
In electromagnetism, to take one example, Eqn. (\ref{4.7}) is
satisfied with $\sigma^{Am}{_{\alpha'}}=0$.  What remains
of Eqn. (\ref{4.6}) in this example is just the 
vanishing of the divergence of the electric charge-current.

Failure of integrability would mean that we have somehow failed
to include in (\ref{2.1}) all the relevant conditions on the first
derivative of the cross-section.  The standard procedure, in
such circumstances, is, first, to enlarge Eqn. (\ref{2.1})
to encompass the additional conditions on
$(\nabla\phi)_m{^{\alpha}}$.  Then look for any
additional constraints arising from this enlargement, and
if any of these fail to be integrable, enlarge
Eqn. (\ref{2.1}) further, etc.  Unfortunately,
it is not clear, in the present general context, how to
implement this procedure.  How do you ``enlarge" a system,
(\ref{2.1}), linear in $(\nabla\phi)_m{^{\alpha}}$, to
encompass a quadratic relation (\ref{4.5})?

Nowhere in this section so far have we introduced a hyperbolization.
It is perhaps striking that so much of the subject of
constraints can be carried out at this level, for it is
largely in their interaction with hyperbolizations that
constraints come to the fore.  We turn now to this interaction.
Fix, therefore, a hyperbolization, $h^A{_{\alpha'}}$, for
Eqn. (\ref{2.1}).

Let $c^{An}$ be a constraint.  Then Eqn. (\ref{4.3}) holds for
the restriction, $\phi_0=\phi |{_S}$, of any solution,
$\phi_0$, of Eqn. (\ref{2.1}) to any three-dimensional
submanifold, $S$, of $M$.  So, in particular, this equation
holds when $(S, \phi_0)$ are initial data, i.e., when the
normal $n_a$ to $S$ at each point $\kappa$ lies in the cone
$s_{\phi_0(x)}$.  Thus, given initial data, $(S,\phi_0)$, we
have no hope of finding a corresponding solution of Eqn.
(\ref{2.1}) unless those data satisfy Eqn. (\ref{4.3}) for
every constraint $c^{An}$.  Eqns. (\ref{4.3}) become 
constraint equations on initial data.

Next, fix $\kappa\in b$ and $n_a\in s_{\kappa}$.  Then, we claim,
for any constraint $c^{An}$ and any (vertical) vector
$v^{\alpha'}$, we can have $n_ac^{Aa}=v^{\alpha'}h^A{_{\alpha'}}$
only if each side is zero.  Indeed, this equality implies
$(n_ac^{Aa})k_A{^m}{_{\beta'}}v^{\beta'}n_m=(v^{\alpha'}h^A{_{\alpha'}}
)k_A{^m}{_{\beta'}}v^{\beta'}n_m$.  But the left side vanishes
(by definition of a constraint), while vanishing of the right
side implies $v^{\alpha'}=0$ (by $n_a\in s_{\kappa}$).  What we
have shown, in other words, is that the subspace of vectors
of the form $n_ac^{Aa}$ with $c^{Aa}$ a constraint, and that of
vectors of the form $v^{\alpha'}h^A{_{\alpha'}}$ with
$v^{\alpha'}$ vertical, have only the zero vector in common.
But this implies that the left side of Eqn. (\ref{4.4})
is less than or equal to the right side.  That is, in the
presence of a hyperbolization, ``half" of Eqn. (\ref{4.4})
is automatic.  Now suppose that the constraints are
complete, i.e., that the full equality (\ref{4.4}) holds.  
It then follows that our two independent subspaces---
$\{ n_ac^{Aa}\}$ and $\{ v^{\alpha'}h^A{_{\alpha'}} \}$---in
fact span the space of all vectors of the form
$\sigma^A$.  What this means, in geometrical terms,
is that the ``constraint components" of Eqn. (\ref{2.1})---the
results of contracting it with vectors of the form
$n_ac^{Aa}$---and the ``dynamical components" of Eqn.
(\ref{2.1})---the result of contracting it with
$h^A{_{\alpha'}}$---together comprise the whole of
Eqn. (\ref{2.1}).  Completeness, in the presence of a
hyperbolization, means the absence of any ``stray equations"
in (\ref{2.1}).

Finally, we return to the issue, raised in the previous
section, of when there is an initial-value formulation
for the full equation (\ref{2.1}).  Fix a hyperbolization
$h^A{_{\alpha'}}$ for this equation, and suppose that its
constraints are both complete and integrable.  Let $(S,\phi_0)$
be initial data, and suppose that these data satisfy all
the constraint equations of the type (\ref{4.3}) (for if
not, then there is certainly no evolution
of these data).  By the general existence-uniqueness theorem
(Sect. 3 and Appendix B), there exists a solution,
$U\stackrel{\phi}{\rightarrow} b$, of the evolution Eqn.
(\ref{3.3}), with $\phi | _S = \phi_0$, where $U$
is an appropriate neighborhood of the three-dimensional
submanifold $S$ of $M$.  We now claim that, under
certain conditions, this cross-section $\phi$ satisfies our
full system, (\ref{2.1}), of partial differential equations.
It is convenient, for purposes of this paragraph, to introduce
upper-case Greek indices to lie in the vector space of
constraints; so, in this notation, we have a single
constraint tensor, $c^{Aa}{_{\Gamma}}$. Denote the left
side of Eqn. (\ref{2.1}), evaluated on the cross-section
$\phi$, by $I_A$.  Thus, we have $h^A{_{\alpha'}}I_A =0$
everywhere in $U$ (by (\ref{3.3})), and $I_A=0$ 
on $S$ (by completeness); and we wish to show $I_A=0$
everywhere in $U$.  To this end, consider the expression
\begin{equation}
(c^{Am}{_{\Gamma}}+\sigma^{m\alpha'}{_{\Gamma}}h^A
{_{\alpha'}})\nabla_m I_A, 
\label{4.9}\end{equation}
where $\sigma^{m\alpha'}{_{\Gamma}}$ is any field on $b$.
We claim that this expression is, everywhere in $U$, a
multiple of $I_A$.  Indeed, the first term in parentheses
leads to such a multiple since, by integrability,
$c^{Am}{_{\Gamma}}\nabla_m I_A$ is a multiple of $I_A$; and the
second term also leads to such a multiple, using
$h^A{_{\alpha'}} I_A = 0$ and differentiating by parts.
To summarize, we have shown so far that $I_A$ vanishes
on $S$, and satisfies a certain first-order, quasilinear
(in fact, linear) partial differential equation arising
from the expression (\ref{4.9}).  Since this differential
equation clearly has as one solution $I_A=0$, we can conclude
that $I_A=0$ in a neighborhood of $S$ if we can show
local uniqueness of its solutions.

The most direct way to prove local uniquenss of solutions of
a partial differential equation is to show that it
admits a
hyperbolization. In the present instance, tensor
$h^{A\Gamma}$ is a hyperbolization of the differential
equation resulting from (\ref{4.9})---for some
choice of $\sigma^{m\alpha'}{_A}$---provided
$h^{A\Gamma}$ has the following property: 
The expression $h^{A\Gamma}c^{Bm}{_{\Gamma}}v_Aw_B$, for all
$v,w$ with $v_A h^A{_{\alpha'}}=w_A h^A{_{\alpha'}}=0$,
is symmetric under interchange of $v$ and $w$, and,
contracted with some $n_m$, is positive-definite.  When---in terms
of the original $k_A{^m}{_{\alpha'}}$ and its hyperbolization
$h^A{_{\alpha'}}$---does such a hyperbolization $h^{A\Gamma}$ exist?
In physical examples---e.g., in electromagnetism---such an
$h^{A\Gamma}$ does indeed exist, and so we have in these
examples uniqueness of solutions of the equation resulting from 
(\ref{4.9}), and so an initial-value formulation for the full system
(\ref{2.1}).  Is there any simple, reasonably general, 
condition on $k_A{^m}{_{\alpha'}}, h^A{_{\alpha'}}$ that 
will guarantee existence of a hyperbolization $h^{A\Gamma}$?  
Are there interesting cases in which uniqueness of
solutions of the differential equation arising from
(\ref{4.9}) must be shown by some other method?

\section{The Combined System---Diffeomorphisms}

In the preceeding sections, we have been analyzing the
structure of the partial differential equation describing
a single physical system.  This analysis was to be applied
separately to the electromagnetic field, a perfect
fluid, or whatever.  However, in the real world, all
these systems coexist on $M$, normally in 
interaction with each other.  We now consider the
system that results from combining all these subsystems.

Once again, we have a smooth fibre bundle,
$B\stackrel{\Pi}{\rightarrow}M$, over the four-dimensional
space-time manifold $M$.  Now, however, the fibre in
$B$ over $x\in M$ represents the possible values at
$x$ of all possible physical fields in the
universe.  Thus, this fibre would include an
antisymmetric tensor (for electromagnetism), a Lorentz
metric and derivative operator (for gravity), two
scalar and one vector field (for a perfect fluid), etc.
Note that we are implicitly assuming that the space that
results from combining these fields is finite-dimensional,
and that we are somehow capable of ``finding" it.
One or both of these assumptions may be incorrect.  In
any case, we imagine that we have constructed such a
bundle.  Again, a cross-section, $M\stackrel{\Phi}{\rightarrow}B$,
of $B$ represents an assignment of a complete physical
state (of everything) to each point of space-time,
i.e., a statement of the entire dynamics of the universe.
And, again, we impose on such cross-sections the general
first-order, quasilinear partial differential equation,
\begin{equation}
K_A{^m}{_{\alpha}}(\nabla\Phi)_m{^{\alpha}}
+J_A = 0.
\label{5.1}
\end{equation} 

In the analogous equation for a single system, (\ref{2.1}),
we allowed the coefficients, $k_A{^m}{_{\alpha}}$ and
$j_A$, to be arbitrary (smooth) fields on the bundle
manifold $b$.  That is, we allowed these coefficients
to depend on both ``the point of the space-time manifold
M and the value of the field assigned to that point".
But, in the context of this combined system, an explicit
dependence of the coefficients, $K_A{^m}{_{\alpha}}$
and $J_A$, on the point of $M$ is, we suggest, inappropriate.
After all, we identify the points of the manifold $M$,
not by somehow ``perceiving them directly",
but rather more indirectly, by observing the various
physical fields on $M$.  So, for instance, the 
physical distinction between two points, $x$ and
$y$, of $M$ rests on the difference between the
values of some physical field at $x$ and at $y$.
(This issue did not arise in the context of a single
system, for there $x$-dependence of $k$ and $j$ could
arise through other physical fields, not included in
the dynamics of (\ref{2.1}).)  In any case, we expect
that the coefficients of $K$ and $J$ in (\ref{5.1})
will depend explicitly only on the fibres of $B$,
with any dependence on the point of $M$ arising only implicitly
through the cross-section.  Unfortunately, this 
expectation---at least, as it is stated above---does
not make mathematical sense!  The problem is that our 
fibre bundles are not naturally products, and so there is no
such thing as ``a function only of the fibre-variables,
independent of the base-space variables".  We must
therefore proceed in a different way. 

We now demand that, as part of the physical content
of the bundle $B$, there be given on it the following
additional structure.  To each diffeomorphism $\cal D$
on the manifold $M$, there is to be 
assigned\footnote{This demand rules out most gauge
theories other than electromagnetism (Appendix A).
What happens in these theories is that to each 
diffeomorphism on $M$ is assigned a number of liftings
to $B$.  Indeed, the collection of all liftings assigned
to the identity diffeomorphism on $M$ is
called the gauge group.  Presumably, much of what follows
could be generalized to include such gauge theories.}
a lifting of it to a diffeomorphism, $\hat{\cal D}$,
on the manifold $B$.   By ``lifting", we mean that 
$\hat{\cal D}$ must satisfy
$\Pi\circ\hat{\cal D} = {\cal D}\circ\Pi$, i.e.,
that $\hat{\cal D}$ must take entire fibres to
fibres, such that the induced diffeomorphism on $M$
is precisely the original $\cal D$.  We further
require of these liftings that they respect the
group structure of the $M$-diffeomorphisms, i.e.,
that $(\widehat{{\rm id}_M})=({\rm id}_B)$ and 
$(\widehat{{\cal D}\circ{\cal D}'}) = \hat{\cal D}\circ
\hat{{\cal D}'}$.  In short, we must specify how the
physical fields ``transform" under diffeomorphisms
on $M$.  In the examples of Appendix A, the
fibres consist of tensors, spinors, derivative
operators, etc., and on such geometrical objects there
is a natural action of $M$-diffeomorphisms.
  Indeed, we claim that
it is this ``transformation behavior" that endows such 
fields with a geometrical content
in terms of $M$.  For instance, given a point of
the tangent bundle of $M$, the direction in $M$ in
which the vector ``points" can be read out from the
$M$-diffeomorphisms whose lifts leave this point invariant.
Consider now the bundle $B$ that results from combining all 
the examples of Appendix A.  Lift diffeomorphisms from $M$ to
this $B$ by combining these liftings for all the individual
examples.

We now have the machinery to express the
idea that the coefficients in Eqn. (\ref{5.1}) be
``functions only of the physical fields".  We demand that,
for every diffeomorphism $\cal D$ on $M$, its lifting
$\hat{\cal D}$ leave $K_A{^m}{_{\alpha}}$ and $J_A$
invariant (noting that this makes sense for
fields having indices in both $B$ and $M$), up to gauge.  It
then follows that, for $\Phi$ any cross-section satisfying   
Eqn. (\ref{5.1}) and $\cal D$ any diffeomorphism on
$M$, the ``transformed cross-section",
$\hat{\cal D}\circ\Phi\circ{\cal D}$, is also a solution.

The ``infinitesimal version" (with diffeomorphisms replaced
by vector fields) of all this is the following.  
To each (smooth) vector field
$\xi^a$ on $M$, there is to be assigned a lifting of it to
a vector field, $\hat{\xi}^{\alpha}$, on $B$.  By
``lifting" we mean that 
$\hat{\xi}^{\alpha}(\nabla\Pi)_{\alpha}{^a}
= \xi^a$.  We require of these liftings that they be
linear (i.e., that $(\widehat{c\xi + \eta}) =
c\hat{\xi} + \hat{\eta}$, for c constant), and Lie-bracket
preserving (i.e., that $\widehat{[\xi, \eta]} =
[\hat{\xi}, \hat{\eta}]$).  Invariance of the
coefficients in (\ref{5.1}) under these infinitesimal
diffeomorphisms now becomes\footnote{The infinitesimal version
of the transformation of solutions of (\ref{5.1}) under
diffeomorphisms becomes that, for every $\xi^a$, its
lifting be a linearized solution of (\ref{5.1}).} 
\begin{equation}
{\cal L}_{\hat{\xi}}K_A{^m}{_{\alpha}}=\Lambda_A
{^m}{_b}(\nabla \Pi)_{\alpha}{^b},  
\; \; \; {\cal L}_{\hat{\xi}}J_A =-\Lambda_A{^m}{_m},
\label{5.2}\end{equation}
for some field $\Lambda_A{^m}{_b}$ on $B$, and for every vector field $\xi^a$ on $M$.
We shall further assume that $\hat{\xi}^{\alpha}$
results from $\xi^a$ through the action of some differential
operator\footnote{It is possible that this assumption follows
already from the general properties above of the liftings of
diffeomorphisms.}:
\begin{equation}
\hat{\xi}^{\alpha} = \delta^{{\alpha}'{m_1}\cdots {m_s}}
{_r}\nabla_{m_1}\cdots \nabla_{m_s} \xi^r+ \cdots.  
\label{5.3} \end{equation}
In (\ref{5.3}), we have written out only the highest-order
term.  Its coefficient, $\delta^{\alpha' {m_1}\cdots {m_s}}{_r} 
=\delta^{\alpha'(m_1\cdots m_s)}{_r}$,
is some smooth field on $B$, independent of the
derivative operator employed in (\ref{5.3}).  Note
that the index ``$\alpha$" of $\delta$ is vertical, as
follows from the definition of a lifting.  As an example, consider
the system resulting from combining all the examples
of Appendix A.  In this case the highest order appearing
in the expression, (\ref{5.3}), for $\hat{\xi}^{\alpha}$ is
$s=2$, and this order occurs only for the derivative operator
of general relativity.  A typical vertical vector 
in the bundle of derivative operators is given by
$\delta\phi^{\alpha'}=\delta\Gamma^a{_{bc}}$ (Appendix A). 
Then the $\delta$ of (\ref{5.3}) 
becomes 
\begin{equation}
\delta^{\alpha'{m_1 m_2}}{_r} = \delta^a{_r}
\delta^{m_1}{_{(b}}\delta^{m_2}{_{c)}},
\label{5.4} \end{equation}
reflecting the action of a Lie derivative on a derivative
operator.

The equation, (\ref{5.1}), for the combined system can never
admit any hyperbolization.  To see this, first note that,
given any three-dimensional submanifold $S$ of $M$,
there always exists a diffeomorphism ${\cal D}$
on $M$ that is the identity in an arbitrarily
small neighborhood of $S$, but not outside that 
neighborhood.  But now, were there a hyperbolization,
then the transformation of solutions $\Phi$ of
(\ref{5.1}) by such a diffeomorphism would violate
the uniqueness theorem for hyperbolic systems (Sect 3
and Appendix B).  Another way to see this is to note that
Eqns. (\ref{5.2}) and (\ref{5.3}) together imply
\begin{equation}
K_{A}{^{(m}}{_{\alpha}}\delta^{|\alpha'|
m_1\cdots m_s)}{_r} = 0,
\label{5.5} \end{equation}
which implies in turn that, for any $n_m$,
the tensor $K_A{^m}{_{\alpha'}}n_m$ is
annihilated on contraction with
$\delta^{\alpha'm_1 \cdots m_s}{_r}n_{m_1} \cdots n_{m_s}$.
But a nonzero vertical vector annihilating $K_A{^m}{_{\alpha'}}n_m$
precludes a hyperbolization.

How, in light of this observation, are we ever to recover an
initial-value formulation in physics (i.e., in
(\ref{5.1}))?  The answer is that we must formulate a
modified version of the initial-value formulation,
in which, given suitable initial data, solutions of
(\ref{5.1}) will always exist, but will be unique
only up to the diffeomorphism freedom.  We now
describe one scheme (suggested by general
relativity) to implement this program.  There may
well be others.  The idea is to supplement
Eqn. (\ref{5.1}) on the cross-section $\Phi$ with a second system of equations,
 of the form
\begin{equation}
\nu_{a\alpha}(\nabla\Phi)_b{^{\alpha}}+
\nu_{ab} = 0,
\label{5.6} \end{equation}
where $\nu_{a\alpha}$ and $\nu_{ab}$ are smooth fields
on $B$.  We wish to so arrange matters that
i) Eqns. (\ref{5.1}) and (\ref{5.6}), taken
together on $B$, have an initial-value formulation,
and ii) Eqn. (\ref{5.6}) can always be achieved, in
a suitable sense, via the diffeomorphism freedom.

We first consider the issue of an initial-value
formulation for Eqns. (\ref{5.1}), (\ref{5.6}). 
A hyperbolization for the system (\ref{5.1}),
(\ref{5.6}) consists of fields $H^A{_{\alpha'}}$ and
$I_{\alpha'}{^{ma}}$ on $B$ such that the expression
\begin{equation}
H^A{_{\alpha'}}K_A{^m}{_{\beta'}} 
+ I_{\alpha'}{^{ma}} \nu_{a\beta'}
\label{5.7} \end{equation} 
is symmetric in indices $\alpha'$, $\beta'$, and
is positive-definite on contraction with some covector
$n_m$ at each point.  What of the constraints?  Let us 
demand that $\nu_{a\alpha}$ have rank four (i.e., that
$\xi^a\nu_{a\alpha} = 0$ implies $\xi^a = 0$).  Then
(\ref{5.6}) represents a sixteen-dimensional vector space
of additional equations on $\Phi$.  The general
constraint for Eqn. (\ref{5.6}) is given by
$c^{An}= x^{abn}$, with $x^{abn}=x^{a[bn]}$.  Thus,
the dimension of the vector space of constraints
is twenty-four, while, for any fixed $n_n$, the
dimension of the space of all $c^{An}n_n$ is just
twelve.  Now consider Eqn. (\ref{4.4}), the condition 
for completeness of the constraints.  As we have
just seen, the act of supplementing Eqn. (\ref{5.1})
by (\ref{5.6}) increases the first term in
Eqn. (\ref{4.4}) by twelve, the second term by
zero, and the third term by sixteen.  Thus, in
order that the system (\ref{5.1}), (\ref{5.6})
be complete, the original system, (\ref{5.1}), must
have yielded a left side of Eqn. (\ref{4.4}) exceeding
the right side by exactly four.  Finally, integrability of the
constraints arising from Eqn. (\ref{5.6}) yields a system of
partial differential equations on the
coefficients $\nu_{a\alpha}$ and $\nu_{ab}$.
The geometrical content of these equations is that
(\ref{5.6}) be precisely the requirement that
the cross-section $\Phi$ lie within a certain submanifold
$V$ of $B$.  This $V$ has codimension four 
(i.e., dimension four less than that of $B$),
and its tangent vectors are those $\xi^{\alpha}$
with $\xi^{\alpha}\nu_{a\alpha}=0$.  

We may summarize the discussion above as follows.
Let there be given a field $\nu_{a\alpha'}$ on $B$
satisfying the following three conditions:
i) for some fields $H$ and $I$, the expression (\ref{5.7})
is symmetric, and, contracted with some $n_m$, is positive-definite;
ii) $\nu_{a\alpha'}$, at each point of $B$, has rank
four, so the space of vectors $v^{\alpha'}$ at
each point with $v^{\alpha'}\nu_{a\alpha'}=0$ has
codimension four; and iii) these vector spaces can be
integrated to give submanifolds, of codimension four,
within each fibre. 
 Given such a field
$\nu_{a\alpha'}$, choose any submanifold $V$ of $B$,
of codimension four, such that $V$ intersects
each fibre of $B$ in one of the submanifolds in
iii) above.  Now restrict the cross-section $\Phi$
to this submanifold $V$, yielding an equation
of the form (\ref{5.6}). 
  By construction, the system
(\ref{5.1}), (\ref{5.6}) admits an initial-value
formulation.

We next turn to the issue of ``achieving" Eqn. (\ref{5.6})
via diffeomorphisms. 
Let $\Phi$ be any solution of Eqn. (\ref{5.1}), and $V$
any submanifold of $B$, as described above, such that,
over some three-submanifold $S$ of $M$, the cross-section
$\Phi$ lies within $V$.  We wish to find a diffeomorphism
on $M$ that sends the entire cross-section $\Phi$ to
lie within the submanifold $V$.  We may choose our
diffeomorphism, together with its first ($s-1$) derivatives,
to be the identity on $S$, but it must begin to differ
from the identity on evolution off $S$.  We shall be able
continually to adjust $\Phi$, via a diffeomorphism, to lie
within $V$ provided we can generate, via (\ref{5.3}), any
$\xi^{\alpha'}$ transverse to $V$.
 But this is precisely the statement
that the operator
\begin{equation}
\nu_{a\alpha'}\delta^{\alpha'm_1 \cdots m_s}{_r}
\nabla_{m_a}\cdots \nabla_{m_s} \xi^r
\label{5.8} \end{equation}
be hyperbolic.  By ``hyperbolic" here, we mean that the
system that results from introducing as auxiliary
fields the first ($s-1$) derivatives of $\xi^a$ admits
a hyperbolization in the sense of Sect. 3. 
See the discussion at the end of
Appendix A.

To summarize, we may recover an initial-value formulation\footnote
{There is a somewhat more elegant, if less accessible, way to
formulate this.  Consider $s=2$ in (\ref{5.3}).  Modify
the bundle $B$ to include, in each fibre, a copy of the
twenty-dimensional manifold of all tensors $\rho_a{^b}$ at
all points of $M$.  Think of a cross-section of this modified
bundle as including a diffeomorphism on $M$ and a
``candidate" ($\rho_a{^b}$) for the derivative of this
diffeomorphism.  Eqns. (\ref{5.1}) and (\ref{5.6}) can be
combined as a hyperbolic system on this modified bundle.
The ``unfolding of the diffeomorphism" is then already
incorporated in the modified bundle.} for Eqn. (\ref{5.1}) provided
we can find a field $\nu_{a\alpha'}$ on $B$ satisfying
conditions i)-iii) above, together with:  iv) the
operator in (\ref{5.8}) is hyperbolic.  As an example
(in fact, {\em the} example) of this scheme, consider 
 Einstein's equation (for the
derivative operator) in general relativity\footnote{For
recent work on the initial-value formulation of Einstein's
equation, see, e.g., S. Fritelli, O. Reula, {\em Comm. Math.
Phys. 266}, 221 (l994); Y. Choquet-Bruhat, J. W. York,
{\em CR Acad Sci, Paris} (to appear).}.  In this example 
$\nu_{a\alpha'}$ is given by
\begin{equation}
\nu_{a\alpha'}\delta\phi^{\alpha'}=
g_{ab}g^{cd}\delta\Gamma^b{_{cd}}.
\label{5.9} \end{equation}
This $\nu$ indeed satisfies the four conditions listed
above.  First note that, for Einstein's equation, the
left side of (\ref{4.4}) does indeed exceed the
right side by four. 
The
$H^A{_\alpha'}$ of Eqn. (\ref{5.6}) is given by
\begin{eqnarray} 
H^A{_{\alpha'}}\delta\phi^{\alpha'}=&
(s^{cdrs}\delta\Gamma^{[a}{_{rs}}t^{b]}+2g^{(d|[a}
s^{b]|c)rs}\delta\Gamma^p{_{rs}}g_{pq}t^q,\nonumber\\
&-\frac{1}{2} s^{abcd}\delta\Gamma^r{_{cd}}g_{rs}t^s),
\label{5.10} \end{eqnarray}
for any $t^a$ timelike, and $s^{abcd} = s^{(cd)(ab)}$ 
positive-definite in its two index pairs.  Using
(\ref{5.4}) and (\ref{5.9}), the operator
of (\ref{5.8}) is just the wave operator.
Note that the field $\nu_{a\alpha'}$ of (\ref{5.9})
is diffeomorphism-invariant.  Thus, the breaking of the
diffeomorphism invariance takes place solely through
the choice of $\nu_{a\alpha}$, i.e., the choice of submanifold $V$.  Are there any other
$\nu_{a\alpha'}$'s that work for general relativity?

The diffeomorphisms, of course, act simultaneously on all the fields
in the combined bundle $B$.  What, then, is the
feature that singles out the field
$\nabla_a$ of general relativity (as opposed, say,
to the field $F_{ab}$ of electromagnetism) to
be the one for which the diffeomorphisms are taken to be the ``gauge"?
The answer, we suggest, is the following.  Let $n_m$ be any
covector in $M$ which, contracted into the expression
(\ref{5.7}), yields a positive-definite quadratic form.
Now apply this quadratic form to the vertical vector
\begin{equation}
v^{\alpha'}=\delta^{\alpha'm_1\cdots m_s}{_a} 
n_{m_1}\cdots n_{m_s}\xi^a.
\label{5.12}\end{equation}
Then the first term arising from (\ref{5.7}) vanishes, by
Eqn. (\ref{5.5}), and so, by positive-definiteness,
we must have $\nu_{a\alpha'}v^{\alpha'}$ nonzero.
We may restate this observation as follows:  That field whose
diffeomorphism-behavior involves the highest number of
derivatives of $\xi^a$ in Eqn. (\ref{5.3}) (i.e., that is
represented by vertical vectors of the form (\ref{5.12}))
is also the field restricted by the gauge-fixing
equation, (\ref{5.6}) (i.e., that corresponds to vertical
vectors not annihilating $\nu_{a\alpha'}$).  In the
case of the combined bundle $B$ resulting from the examples
of Appendix A, the highest number of derivatives of
$\xi^a$ arising from diffeomorphism-behavior is $s=2$,
and the field having this behavior is, of course, the
derivative operator $\nabla_a$.  In this manner, the
derivative operator of general relativity acquires the
diffeomorphisms as its gauge group.  These remarks have
the following curious consequence.  Suppose that, at some
time in the future, there were introduced a new physical
field, having order $s=3$ in Eqn. (\ref{5.3}).  Then,
apparently, that new field would take the diffeomorphisms
as its gauge group; leaving no ``gauge freedom" in
general relativity.

\section{Physical Systems---Interactions}

We adopt the view that the combined bundle $B$, with
its partial differential equation (\ref{5.1}) and
action of the diffeomorphism group, comprises
all there is in the (nonquantum) physical world.
But, by contrast, we do not view our world as such
a single entity.  Rather, it appears to be divided
into various ``physical systems".
For example, one such system is comprised of the electromagnetic field $F_{ab}$ alone;
another, of the four-velocity $u^a$, mass density $\rho$,
and particle-number density $n$ for a perfect fluid.
 We do not, e.g., organize these
four fields into one system $(F_{ab}, n)$, and another
$(u^a,\rho)$.  We then think of these individual systems 
as ``interacting" with each other.  Interactions take place
on a number different of levels.  Consider, for example, the
electromagnetic field $F_{ab}$.  In the absence of
a space-time metric, there is no natural choice
for the electromagnetic field tensor ($F_{ab}$, or
$F^{ab}$, or some density?); and, even if some
such choice were made, there is no way to write
down Maxwell's equations.  (Note that neither of 
these assertions is true with the roles of the metric and
electromagnetic field reversed.)  This is one level of
interaction.  On a different level is the interaction of
a charged fluid on
the electromagnetic field  through
the appearance of the the fluid charge-current
in Maxwell's equations. 
These structural features of the world---the
notion of physical systems and their various levels
of interaction---must
somehow be ``derived" from Eqn (\ref{5.1}) on $B$---at
least, if our view of the primacy of (\ref{5.1})
is to be maintained. 
How all this comes about is the subject of
this section.

Let $b\stackrel{\pi}{\rightarrow}M$ be a fibre bundle.  By a
{\em quotient bundle} of $b$, we mean a smooth manifold
$\hat{b}$, together with smooth mappings
$b\stackrel{\zeta}{\rightarrow}\hat{b}\stackrel{\hat{\pi}}
{\rightarrow}M$, such that i) $\hat{\pi}\circ\zeta=\pi$,
and ii) $\hat{b}\stackrel{\hat{\pi}}{\rightarrow}M$ is
a fibre bundle (over $M$), and $b\stackrel{\zeta}{\rightarrow}
\hat{b}$ is a fibre bundle (over $\hat{b}$).  Thus, a
quotient ``inserts a manifold $\hat{b}$ between $b$
and $M$, in such a way that there is created a fibre
bundle on each side of $\hat{b}$".  The following
example will illustrate both the mathematical structure
and the types of applications we have in mind.  Let
$b$ be the bundle whose fibre over $x\in M$ consists
of pairs $(g_{ab},F_{ab})$, where $g_{ab}$ is a 
Lorentz-signature metric at $x$, and $F_{ab}$ 
an antisymmetric tensor (the electromagnetic field).
Now let $\hat{b}\stackrel{\hat{\pi}}{\rightarrow}M$ be the bundle
whose fibres include only the Lorentz metrics, and let
${b}\stackrel{\zeta}{\rightarrow}\hat{b}$ be the
mapping that ``ignores $F_{ab}$".  This $b\stackrel{\zeta}
{\rightarrow}\hat{b}\stackrel{\hat{\pi}}{\rightarrow}M$,
we claim, is a quotient bundle.  Furthermore, it reflects
the natural relationship between $F$ and $g$, i.e.,
that it is meaningful to discard $F_{ab}$ while
retaining $g_{ab}$, but not to discard $g_{ab}$ while
retaining $F_{ab}$.  

Let $b\stackrel{\zeta}{\rightarrow}\hat{b}\stackrel
{\hat{\pi}}{\rightarrow}M$ be a quotient bundle. 
Fix any cross-section, 
$M\stackrel{\hat{\phi}}{\rightarrow}\hat{b}$, of $\hat{b}$.
We now construct, using this $\hat{\phi}$, a new
fibre bundle, $\check{b}\stackrel{\check{\pi}}{\rightarrow}M$, as
follows.  For the manifold $\check{b}$, we take the submanifold
$\zeta^{-1}[\hat{\phi}[M]]$ of $b$ (i.e., the set
of points of $b$ lying above the fixed cross-section
$\hat{\phi}$ of $\hat{b}$); and for the mapping
$\check{\pi}$ we take the restriction to the submanifold
$\check{b}$ of the projection $\pi$.  The bundles
$\hat{b}\stackrel{\hat{\pi}}{\rightarrow}M$ and
$\check{b}\stackrel{\check{\pi}}{\rightarrow}M$ represent a kind
of ``splitting" of the original bundle 
$b\stackrel{\pi}{\rightarrow}M$.  For instance, we have
(dim fibre $\hat{b}$) + (dim fibre $\check{b}$) =
(dim fibre $b$).  Every cross-section, $\phi$, of $b$
yields both a cross-section of $\hat{b}$ (namely,
$\hat{\phi} = \zeta  \circ\phi$), and a cross-section
of $\check{b}$ (namely, $\phi$).  And, conversely,
cross-sections of $\hat{b}$ and $\check{b}$ combine
to form a cross-section of $b$.
 But---and
this is the key point of the construction---the
bundle $\check{b}$ requires for its very existence a given
cross-section of $\hat{b}$.  We illustrate this construction
with our earlier example (with $b$-fibres $(g_{ab}, F_{ab})$,
and $\hat{b}$-fibres $(g_{ab})$).  Fix a cross-section
of $\hat{b}$ i.e., fix a metric field $\tilde{g}_{ab}$ on $M$.
  Then the submanifold
$\check{b}$ of $b$ consists of all $(x, \tilde{g}_{ab},
F_{ab})$.  That is, the metric at each $x\in M$ is required to be
the fixed $\tilde{g}_{ab}$ there, while the electromagnetic
field $F_{ab}$ at $x$ remains arbitrary.  So, $\check{b}$
in this example is the bundle over $M$ of electromagnetic
fields, in the presence of the fixed background
metric $\tilde{g}_{ab}$.  Clearly, a cross-section of the
bundle $\hat{b}$ (metric field $\tilde{g}_{ab}$),
together with a cross-section of $\check{b}$ (an electromagnetic field in the presence
of background metric $\tilde{g}_{ab}$),
yield a cross-section of
the original bundle $b$; and conversely.

The notion of a quotient bundle captures the idea of one set 
of physical fields serving as the kinematical background 
for another.  Thus, the metric serves as the kinematical background
field for the electromagnetic field; and the metric and
electromagnetic fields together serve as the kinematical
background for a charged scalar field.  We turn now
from kinematics to dynamics.  

Fix a fibre bundle $b\stackrel{\pi}{\rightarrow}M$, and
a quotient bundle thereof, $b\stackrel{\zeta}{\rightarrow}
\hat{b}\stackrel{\hat{\pi}}{\rightarrow}M$.  Then, as
we have just seen, the cross-sections of $b$ 
``split", in the sense that specifying a cross-section
$\phi$ of $b$ is equivalent to specifying a cross-section
$\hat{\phi}$ of $\hat{b}$, together with a cross-section
$\check{\phi}$ of the bundle $\check{b}$ derived from $\hat{\phi}$.  Next,
let there be specified a system of quasilinear, first-order
partial differential equations, (\ref{2.1}), on cross-sections
$\phi$ of $b$.  When does this equation also split
into separate equations on the cross-sections, 
$\hat{\phi}$ and $\check{\phi}$, that comprise $\phi$?
We are here concerned only with the first term in
(\ref{2.1})---the dynamical part of the differential
equation.  The remainder will be discussed shortly.
So, fix a field $k_A{^m}{_{\alpha'}}$ on $b$:  We wish
to split it into corresponding fields on $\hat{b}$ and
$\check{b}$.  There is a natural choice for the field on $\check{b}$,
namely the restriction to the submanifold $\check{b}$ of $b$
of the given field $k_A{^m}{_{\alpha'}}$ on $b$.
On taking this restriction, only some of the 
components of $k_A{^m}{_{\alpha'}}$ survive, namely, 
those represented by contraction of $k_A{^m}{_{\alpha'}}$
with vectors $v^{\alpha'}$ tangent to the submanifold
$\check{b}$.  The remaining components---those lost under this
restriction---must now be recovered from a suitable field,
$\hat{k}_{\hat{A}}{^m}{_{\hat{\alpha}'}}$ on $\hat{b}$.  
This recovery will occur provided i) the pullback
of $\hat{k}_{\hat{A}}{^m}{_{\hat{\alpha}'}}$ from
$\hat{b}$ to $b$ is a linear combination of
$k_A{^m}{_{\alpha'}}$ on $b$, and ii) this pullback,
together with the restriction of $k_A{^m}{_{\alpha'}}$
to $\check{b}$, exhausts $k_A{^m}{_{\alpha'}}$.  The first
condition means, in more detail, that, for
some field $\mu_{\hat{A}}{^A}$ on $b$, we have
\begin{equation}
(\nabla\zeta)_{\alpha'}{^{\hat{\alpha}'}}k_{\hat{A}}{^m}
{_{\hat{\alpha}'}}= \mu_{\hat{A}}{^A}k_A{^m}{_{\alpha'}}, 
\label{6.1} \end{equation}
where the right side is evaluated at $\kappa\in b$, the
left side at $\zeta(\kappa)\in \hat{b}$.  The left
side of (\ref{6.1}) is the pullback of
$\hat{k}_{\hat{A}}{^m}{_{\hat{\alpha}'}}$ from $\hat{b}$
to $b$ via the mapping $\zeta$; the right side, some
linear combination (with coefficients $\mu$)
of $k_A{^m}{_{\alpha'}}$.  This condition guarantees
that the dynamical part of the differential equation to be imposed on
$\hat{b}$ will come from the dynamical part of the differential
equation originally given on $b$. 
The second condition means, in more detail, that, for
any vector $\sigma^A$ such that the restriction of
$\sigma^Ak_A{^m}{_{\alpha'}}$ to $\check{b}$ vanishes,
we have $\sigma^A=\tau^{\hat{A}}\mu_{\hat{A}}{^A}$ for some
$\tau^{\hat{A}}$.  In other words, what is lost on
restriction (to $\check{b}$) must be regained via the pullback
(from $\hat{b}$).  Note that such a $\hat{k}$ on $\hat{b}$,
if it exists, is unique.  

Given fibre bundle $b\stackrel{\pi}{\rightarrow}M$, and
field $k_A{^m}{_{\alpha'}}$ on $b$, by a {\em reduction}
of this we mean a quotient bundle $\hat{b}$, with field
$\hat{k}_{\hat{A}}{^m}{_{\hat{\alpha}'}}$ on $\hat{b}$, 
satisfying the two conditions above (in the case of
ii), for every cross-section $\hat{\phi}$).  To illustrate
this definition, we return again to our earlier example.  On the
metric-electromagnetic bundle $b$ above, introduce the
field $k_k{^m}{_{\alpha'}}$ arising from the metric-Maxwell
equations:
\begin{equation}
\nabla_ag_{bc}=0,\; \; \; \nabla^bF_{ab}=0,\; \; \;
\nabla_{[a}F_{bc]}=0.  
\label{6.2} \end{equation}
Let $\hat{b}$ be the quotient bundle above (in which only
the metric $g_{ab}$ is retained).  On $\hat{b}$, introduce the
field $\hat{k}_{\hat{A}}{^m}{_{\hat{\alpha}'}}$ arising
from the first equation in (\ref{6.2}).  (Here, we are
making essential use of the fact that no electromagnetic
field appears in this equation.)
This $\hat{b}$,
$\hat{k}_{\hat{A}}{^m}{_{\hat{\alpha}'}}$, we claim,
is a reduction.  Indeed, condition i) follows from the
fact that, whenever cross-section $(g_{ab}, F_{ab})$ of
the bundle $b$ satisfies the full system (\ref{6.2}), then
the corresponding cross-section $g_{ab}$ of $\hat{b}$
satisfies the first of these equations.  Condition ii)
follows from the fact that, given a field $g_{ab}$
(cross-section of $\hat{b}$) satisfying the first
equation, and then an $F_{ab}$ (cross-section of $\check{b}$)
satisfying, in the presence of that $g_{ab}$ as
background, the last two equations, then the full
set, $(g_{ab}, F_{ab})$, satisfies the full system
(\ref{6.2}).  Note, by contrast, that there is no
reduction with the roles of the electromagnetic
field and the metric reversed. 

Let us now return to the combined bundle $B$, with its
partial differential equation (\ref{5.1}).  This 
$(B, K_A{^m}{_{\alpha'}})$ will, of course, have numerous
reductions; and the $(\hat{B}, \hat{K}_{\hat{A}}{^m}{_{\hat
{\alpha}'}})$ that result may have further reductions;
and so on.  Any fields lost (i.e., incorporated 
in bundle $\check{B}$) in such a reduction will
be called a {\em physical system}; while the fields remaining
(in the bundle $\hat{B}$) will be called {\em background fields}
for that physical system.  These definitions, we suggest,
capture the way in which fields are grouped together
in physics, and the sense in which the equations for some
fields require as prerequisites other fields.  For example,
the fields $(u^a,\rho,n)$ for a perfect fluid form a physical
system (but not, e.g., $(\rho, n)$ alone); with background the 
space-time metric.  The charged Dirac field is a physical
system, with background the electromagnetic and metric
fields together; and the electromagnetic field is a
physical system, with background the metric field.  This definition 
also produces a couple of minor surprises.  For the
Klein-Gordon equation, the scalar and vector fields,
$\psi$ and $\psi_a$, form separate physical systems, the
former having no background fields; the latter, just
the metric field.  (Thus, neither of these has the other
as background!)  Similarly, the metric and
derivative operator of general relativity form separate
physical systems, the metric having no background, the
derivative operator the metric as background.  Note, from the examples
of Appendix
A, that the metric is a background for virtually every
physical system (the sole exceptions being those, such as the
Klein-Gordon $\psi$, having such a wide variety of
hyperbolizations that every covector in $M$ lies in the cone
$s_{\kappa}$ for at least one of them).  This feature
presumably reflects the fact that, in order that
there be a hyperbolization for the combined system
(\ref{5.1}), it is not enough merely that there be
a hyperbolization for each individual physical
system making up $B$.  These individual hyperbolizations must also be such that
they have in their various $s_{\kappa}$ a common $n_m$.  In order to achieve
this, the individual physical systems must somehow arrange
to ``communicate" with each other what their dynamics is.  That communication 
takes place by sharing
the space-time metric $g_{ab}$ as a background field.

So far, we have focussed exclusively on the ``dynamical part"
of Eqn. (\ref{5.1})---the $K_A{^m}{_{\alpha'}}$.  We turn now
to the remainder of this equation---the $J_A$.  Roughly
speaking, whenever that portion of Eqn. (\ref{5.1}) that
specifies the dynamics of one physical system has its
$J_A$ depending on the fields of another, then we say that
the second system interacts on the first.  However, we must
exercise some care in formulating this idea precisely.  
For instance, $J_A$ is not gauge-invariant (and, indeed,
can, by a suitable gauge transformation, be made to vanish);
and furthermore it is unclear what ``depends on" is to mean
for fields on a bundle space.

Consider again Eqn. (\ref{2.1}).  Fix a reduction of that system,
so we have a quotient bundle $b\stackrel{\zeta}{\rightarrow}
\hat{b}\stackrel{\hat{\pi}}{\rightarrow}M$, together with
a field $k_A{^m}{_{\alpha'}}$ on $b$, satisfying
conditions i) and ii) above.  Any cross-section,
$\hat{\phi}$, of $\hat{b}$ gives rise to a new bundle,
$\check{b}$; and this $\hat{\phi}$, together with a
cross-section $\check{\phi}$ of $\check{b}$, specifies
a cross-section $\phi$ of the original bundle $b$.
Furthermore, the dynamical term $k_A{^m}{_{\alpha'}}$ on
$b$ splits into corresponding dynamical terms on
$\hat{b}$ and $\check{b}$.  We say that the physical
system represented by cross-sections $\check{\phi}$ of
$\check{b}$ {\em interacts on} the physical system
represented by cross-sections $\hat{\phi}$ of $\hat{b}$
provided there is no similar way to split the entire
partial differential equation (\ref{2.1}) on $b$.  Thus,
in more detail, the $\check{b}$-system interacts on the
$\hat{b}$-system provided there exist no fields
$\hat{k}_{\hat{A}}{^m}{_{\hat{\alpha}}}$, 
$\hat{j}_{\hat{A}}$ on $\hat{b}$ such that
\begin{equation}
(\nabla\zeta)_{\alpha}{^{\hat{\alpha}}}\hat{k}_{\hat{A}}
{^m}{_{\hat{\alpha}}}=\mu_{\hat{A}}{^A}k_A{^m}{_{\alpha}},
\; \; \; \hat{j}_{\hat{A}}= \mu_{\hat{A}}{^A}j_A,
\label{6.3} \end{equation}
up to gauge.  This is to be compared with Eqn. (\ref{6.1}).
We are merely strengthening condition ii) of the definition
of a reduction to require an entire partial differential
equation on $\hat{b}$ whose pullback is a linear
combination of the given partial differential equation
on $b$.  In order words, if we can write the equations on
$\hat{b}$ in such a way that ``no $\check{b}$-fields are
involved", then $\check{b}$ does not interact on $\hat{b}$.

As an example, consider an electromagnetic field and uncharged
perfect fluid, in the presence of a background metric and
derivative operator.  Then the electromagnetic field does
not interact on the perfect fluid:  We can find a reduction
of this system in which the electromagnetic field is
carried in $\check{b}$, the perfect fluid in $\hat{b}$; and
a system of equations on $\hat{b}$ (the perfect-fluid
equations, described by $\hat{k}_{\hat{A}}{^m}{_{\hat{\alpha}}}$
and $\hat{j}_{\hat{A}}$) not involving the electromagnetic
field.  Similarly, the perfect fluid does not interact on the
electromagnetic field.  However, if this is a charged
perfect fluid, then each of these physical systems interacts
on the other (the perfect fluid on the electromagnetic field
through the charge-current term in Maxwell's equations; the
electromagnetic field on the perfect fluid through the
Lorentz-force term in the fluid equations).  As a second
example, note that the derivative operator interacts on virtually
every physical system (through the use of this derivative operator
in writing out field equations); and virtually every
physical system interacts on the derivative operator
(through Einstein's equation).

We remark that the definition is manifestly gauge-invariant.
Note that, as the definition is formulated, ``interacts on"
is not defined at all unless we have an appropriate reduction.
Thus, for example, we are not permitted even to ask whether
the metric interacts on the electromagnetic field.
(Perhaps it would be more natural to extend the definition
so that background fields for a given physical system 
automatically interact on that system.)  Finally, we remark
that ``interact on" need not be reciprocal:  It is possible for
system A to interact on B, but not B on A.  It is not difficult
to construct an example, e.g., with A a Klein-Gordon system and
B a perfect fluid.  Introduce an additional term, involving
the Klein-Gordon fields, on the right side of the equation,
(\ref{A.14}), of fluid particle-number conservation.  
Conservation of fluid stress-energy is not thereby disturbed.
However, I am not aware of any such examples arising naturally
in physics.  Should this observation be elevated to a 
general principle?

We are, in a real sense, finished at this point.  The world
is described---once and for all---by the combined bundle
$B$, with its cross-sections subject to (\ref{5.1}).  In
particular, whatever interactions there are in the world have
already been included in the term $J_A$ of this equation.
However, it is traditional to think of interactions
in a somewhat different way---to think of them as capable
of being ``turned on and off" by some external agency
(presumably, us).  Consider, for example, the 
electromagnetic-charged fluid system.  The fields are $F_{ab}$,
$n$, $\rho$, and $u^a$; and there appears, on the right side
of the Maxwell equation (\ref{A.1}), a term $enu^a$ 
(charge-current), and, on the right side of the fluid
stress-energy conservation equation (\ref{A.12}) a term
$enF^a{_m}u^m$ (Lorentz force).  Here, $e$ is some fixed number
(charge per particle).  In the traditional view for
this particular system, we think of this system as
arising, not full-blown in its final form, but rather
in two distinct steps.  First, introduce the system with
``no interaction" ($e=0$), and then ``turn on the interaction"
by adjusting $e$ to its correct value.

This traditional view may be expressed, in the present general
framework, as follows.  On the combined bundle $B$, there is to
be specified a ``basic version" of the dynamical equations,
(\ref{5.1})---a version in which ``all interactions that can be turned
off have been."  Thus, the ``electromagnetic interaction" has
been turned off in the basic version:  The only physical system
that interacts on the electromagnetic field is the derivative
operator.  The ``gravitational interaction" has been turned off:
No field interacts on the derivative operator.  For more
complicated interactions---e.g., those of contact forces between
materials---it may not always be clear how this ``turning off" is
to be carried out.  The derivative operator generally
survives into the basic version to interact with most
physical systems.  Physically, this phenomenon is a reflection
of the equivalence principle.  Mathematically, it arises
because it is difficult to eliminate the derivative operator
and still lift diffeomorphisms (i.e., ``maintain covariance")
as in Sect. 5. 
This role of the derivative operator is related to the fact
that the behavior of $\nabla_a$ under an infinitesimal
diffeomorphism, given in Eqn. (\ref{5.3}), involves the
second derivative of the vector field $\xi^a$, whereas
other physical fields involve only the first derivative.
To see this, consider a physical field, such as the $F_{ab}$ for
electromagnetism, having $s=1$ in (\ref{5.3}).  Consider
that portion of the combined equation, (\ref{5.1}),
referring to the electromagnetic field, and apply to it the
infinitesimal diffeomorphism generated by $\xi^a$.  Then the
term ``$(\nabla\Phi)_m{^{\alpha}}$" acquires a second
derivative of $\xi^a$.  So, invariance of (\ref{5.1}) under
infinitesimal diffeomorphisms can be maintained only if a
second derivative of $\xi^a$ appears elsewhere in the electromagnetic
portion of the 
equation, i.e., in the term $J_A$.  That is, some physical
field must have, for its transformation behavior 
(\ref{5.3}) under infinitesimal diffeomorphisms, $s=2$.
That field is the derivative operator.

In addition to this basic version of (\ref{5.1}), there is
to be given some ``interaction fields", i.e., some fields
$\delta J_A$ on $B$.  We are free to add, to the left side
of Eqn. (\ref{5.1}), any linear combination, with constant
coefficients (the coupling constants), of these $\delta J_A$.
This is the step of turning on the interactions.  One such
$\delta J_A$, for instance, is that which inserts the
fluid charge-current $nu^a$ into (\ref{A.1}), and the Lorentz
force $nF^a{_m}u^m$ into (\ref{A.12}).  Another inserts the
trace-reversed combined stress-energies of all physical
systems into\footnote{Note that these individual stress-energies
cannot be inserted one at a time, with separate $\delta J_A$,
for to do so would violate integrability of the constraints
of the combined system.}  Einstein's equation, (\ref{A.21}).  
It is not entirely clear how much of this traditional view
is psychological and how much physical.  One way to argue that it
has physical content might be to produce a general construction,
given only the full combined Eqn. (\ref{5.1}), that
yields the basic version of this equation, as well as
the appropriate $\delta J_A$. 

These $\delta J_A$ do not span, at each point of the bundle
$B$, the space of vectors of type ``$\sigma_A$".  That is,
there are algebraic restrictions on the allowed interactions.
These restrictions appear to be an essential part of the
physical content of the systems under considerations.
Disallowed, for example, are interactions on the
electromagnetic field that represent a magnetic charge-current;
on a perfect fluid that provide a source for particle-number
conservation; and on the metric in (\ref{A.19}).  See
Appendix A for further examples.

The basic version of Eqn. (\ref{5.1}) on $B$ must certainly be
a viable system of equations, and so in particular its
constraints must be complete\footnote{That is, complete in
the sense appropriate to the diffeomorphism freedom inherent in
(\ref{5.1}), namely that the left side of Eqn.
(\ref{4.4}) exceeds the right side by four.} and integrable.
What happens to viability of this system on turning on
some interaction $\delta J_A$?  Completeness of the
constraints will not change, for this involves only the 
dynamical term, that with coefficient $K_A{^m}{_{\alpha'}}$
in (\ref{5.1}).  But integrability could be destroyed
by turning on such an interaction.  Indeed, the 
necessary and sufficient condition that integrability
be retained for fixed constraint $c^{An}$ of Eqn.
(\ref{5.1}) under addition to $J_A$ a term $\delta J_A$
is that, for some $\tau^A$ and $\tau_a{^b}$ with
$\tau_a{^a}=0$, we have
\begin{eqnarray}
&&\nabla_{\alpha}(c^{Am}\delta J^A)=
\sigma^{Am}{_{\alpha}}\delta J_A \nonumber \\
&&\ \ \ \ +\tau^A
(K_A{^m}{_{\alpha}}+ \frac{1}{4}(J_A
+\delta J_A)(\nabla\pi)_{\alpha}{^m})
+\tau^m{_s}(\nabla\pi)_{\alpha}{^s},
\label{6.4}\end{eqnarray}
where $\sigma^{Am}{_{\alpha}}$ is the tensor that appears
in (\ref{4.6}) for the basic version of Eqn. (\ref{5.1}).  The proof of
this assertion is straightforward:  Demand that the
integrability condition, (\ref{4.6}), be satisfied for
both $K_A{^m}{_{\alpha}}$, $J_A$ and $K_A{^m}{_{\alpha}}$,
$J_A+\delta J_A$.  Thus, any viable candidate
$\delta J_A$ for an interaction that can be ``turned on"
must satisfy the condition of (\ref{6.4}).

Eqn. (\ref{6.4}) is apparently a rather severe restriction
on the interactions allowed in nature.  The main reason for this 
is its nonlinearity:  Given $\delta J_A$ and
$\delta J'_A$, for each of which there exist $\tau$'s
satisfying (\ref{6.4}), we have no guarantee that there
exist $\tau$'s that work for their sum.  Here is a physical
example of this behavior.  Let $\delta J_A$ insert a perfect-fluid
stress-energy into Einstein's equation, and $\delta J'_A$
insert an electromagnetic interaction into the perfect-fluid
equations.  Then each of these modifications of (\ref{5.1})
preserves integrability of the gravitational constraint, but
their sum, $\delta J_A + \delta J'_A$, does not.  To
achieve integrability in this example, it is necessary to
add to this sum a term that also inserts the electromagnetic
stress-energy into Einstein's equation.  Indeed, it is
not even obvious, from (\ref{6.4}), that whenever $\delta J_A$
preserves integrability in (\ref{5.1}), then so does
$2\: \delta J_A$!

So, finding a collection of interaction expressions,
$\delta J_A$, that can be turned on in any linear
combination, preserving all the while integrability of
the constraints of (\ref{5.1}) does not appear to be
easy.  So, how does nature accomplish this?  Is there, for
example, some simple, general criterion that can be 
applied to the $\delta J_A$'s to guarantee integrability?

\appendix
\section*{Appendix A---Examples}
{\bf Electromagnetism}

The field is an antisymmetric tensor field, $F_{ab}=F_{[ab]}$, on
$M$, with background metric $g_{ab}$.  The equations are
\begin{equation}
\nabla^bF_{ab} =0,
\label{A.1}\end{equation}
\begin{equation}
\nabla_{[a}F_{bc]}=0.
\label{A.2}\end{equation}

Thus, the fibres of the bundle $b$ are six-dimensional, a typical
vertical vector being given by $\delta\phi^{\alpha'}=\delta F_{ab}
= \delta F_{[ab]}$.  A typical vector in the space of equations is
given by $\sigma^A = (s^a, s^{abc})$, where $s^a$ is a vector
(the coefficient of (\ref{A.1})), and $s^{abc}$ an
antisymmetric third-rank tensor (the coefficient
of (\ref{A.2})).  Thus, the equation-space is eight-dimensional.

The general hyperbolization at a point is given by
\begin{equation}
h^A{_{\alpha'}}\delta\phi^{\alpha'}=(\delta F^a{_m}t^m,\; \; - \frac{3}
{2}  t^{[a}
\delta F^{bc]}),
\label{A.3}\end{equation}
where $t^a$ is an arbitrary timelike vector.  The general constraint
at a point is given by
\begin{equation}
c^{An}=(xg^{an},\; \; y\epsilon^{abcn}), 
\label{A.4}\end{equation}
where $x$ and $y$ are numbers.  Thus, the constraints form a 
two-dimensional vector space, while the space of vectors of the form
$c^{An}n_n$, for fixed $n_n$, is also two-dimensional.  The constraints
are complete and integrable. 

 The allowed interactions are
\begin{equation}
\delta j_A = (j_a,\; \; 0),
\label{A.5}\end{equation}
i.e., no magnetic charge-current is allowed. 
These will preserve integrability provided $\nabla^aj_a =0$.
\\

\noindent {\bf Klein-Gordon}

The fields consist of a scalar field, $\psi$, and a vector
field, $\psi_a$, on $M$, with background metric $g_{ab}$.  The
equations are
\begin{equation}
\nabla_a\psi = \psi_a,
\label{A.6}\end{equation}
\begin{equation}
\nabla_{[a}\psi_{b]}=0,
\label{A.7}\end{equation}
\begin{equation}
\nabla^a\psi_a=0.
\label{A.8}\end{equation}

Thus, the fibres of the bundle $b$ are five-dimensional, a typical
vertical vector being given by $\delta\phi^{\alpha'}=
(\delta\psi, \delta\psi_a)$.  A typical vector in the space of
equations is given by $\sigma^A=(s^a, s^{ab}, s)$ (the
respective coefficients of (\ref{A.6})-(\ref{A.8})), where
$s^{ab}$ is antisymmetric.  Thus, the equation-space is
eleven-dimensional.

The general hyperbolization at a point is given by
\begin{equation}
h^A{_{\alpha'}}\delta\phi^{\alpha'}=(
-w^a\delta\psi,\; \; -t^{[a}\delta\psi^{b]},\; \; \frac{1}{2}
t^a\delta\psi_a),
\label{A.9}\end{equation}
where $t^a$ is any timelike vector (say, future-directed), and $w^a$ any vector 
not past-directed timelike or null.   In order that this hyperbolization be causal,
we must require in addition that $w^a$ be future-directed timelike
or null.   The
general constraint at a point is given by
\begin{equation}
c^{An}=(x^{an},\;  y^{abn},\; 0),
\label{A.10}\end{equation}
where $x^{an}$ and $y^{abn}$ are arbitrary antisymmetric tensors.
Thus, the constraints form a ten-dimensional vector space,
while the space of vectors of the form $c^{An}n_n$, for fixed
$n_n$, is six-dimensional.  The constraints are complete and
integrable. It turns out that there are two separate physical systems
(in the sense of Sect. 6) here.  The first involves the field
$\psi$ alone: The equation is  (\ref{A.6}), the hyperbolizations
those generated
by $w^a$ in Eqn. (\ref{A.9}), and the constraints those generated by
$x^{an}$ in Eqn. (\ref{A.10}).  There are no background fields
for this physical system.  The
other involves the field $\psi_a$: The equations are (\ref{A.7}),
(\ref{A.8}), the hyperbolizations those generated by $t^a$ in
Eqn. (\ref{A.9}), and the constraints those generated by $y^{abn}$
in Eqn. (\ref{A.10}).  The background field is the metric.

The situation regarding allowed interactions is not entirely
clear (reflecting, perhaps, a certain lack of physical context
for this system).  Certainly, one allowed interaction is
\begin{equation}
j_A=(0,\; \; 0,\; \;- m^2\psi),
\label{A.11}\end{equation}
where $m$ is a number.  This results in the massive
Klein-Gordon system.  (Note that the ``tachyon equation"---
the result of letting $m$ be imaginary in (\ref{A.11})---
admits a hyperbolization.)   
 It seems likely that all allowed interactions
have zeros in the first two entries of (\ref{A.11}), for
otherwise it is difficult to preserve integrability.  (Passage
to a charged Klein-Gordon system is not turning on an
interaction in our sense, for it requires an entirely
new bundle $b$.  This example is discussed later in this Appendix.)
Whether there are allowed other interactions of the form (\ref{A.11}), but
with different third entries on the right (for all of which,
incidentally, all the constraints are integrable), is unclear. 
\\
 
\noindent{\bf Perfect Fluid}

The fields consist of two scalar fields, $n$ and $\rho$, and a
unit timelike vector field, $u^a$, on $M$,
with background metric $g_{ab}$.  The equations are
\begin{equation}
(\rho + p)u^m\nabla_mu^a+(g^{ab}+u^au^b)\nabla_bp=0,
\label{A.12}\end{equation}
\begin{equation}
u^m\nabla_m\rho+(\rho+p)\nabla_mu^m=0,
\label{A.13}\end{equation}
\begin{equation}
u^m\nabla_mn+n\nabla_mu^m=0,
\label{A.14}\end{equation}
where $p(n,\rho)$ is some fixed function (the function
of state).  The first two equations are the components of
conservation of $T^{ab} = (\rho+p)u^au^b+pg^{ab}$, the
third conservation of $N^a=nu^a$.

Thus, the fibres of the bundle $b$ are five-dimensional, a
typical vertical vector being given by $\delta\phi^{\alpha'}
=(\delta n,\delta\rho, \delta u^a)$, with $\delta u^a$
(because of unit-ness of $u^a$) orthogonal to $u^a$.  A
typical vector in the space of equations is given by
$\sigma^A=(s_a,s,\hat{s})$ (respective coefficients of
(\ref{A.12})-(\ref{A.14})), with $s_a$ orthogonal
to $u^a$ (reflecting that the left side of Eqn.
(\ref{A.12}) is).  The equation-space is five-dimensional.

This physical system admits no hyperbolization unless
\begin{equation}
\rho+p>0, \; \; \; (\frac{\partial p}{\partial \rho}
+\frac{n}{\rho+p}\frac{\partial p}{\partial n}) 
>0. 
\label{A.16}\end{equation}
Physically, this is the requirement that inertial
mass and sound speed both be positive. 
So, the fibres of the bundle $b$ must be suitably restricted to
achieve (\ref{A.16}) everywhere.   
The most general hyperbolization at a point is then given by
\begin{eqnarray}
h^A{_{\alpha'}}\delta\phi^{\alpha'}=x(((\rho+p)\frac{\partial p}{\partial\rho
}+n\frac{\partial p}{\partial n})\delta u_a,\; \; 
\frac{\partial p}{\partial\rho}\delta p,\; \; \frac{\partial p}{\partial n}
\delta p)\nonumber \\ 
+y(\delta n - \frac{n}{\rho+p}\delta\rho)
(0,\; \; \frac{n}{\rho+p},\; \; -1),
\label{A.15}\end{eqnarray}
where $x$ and $y$ are any numbers with $xy>0$, and where we have
set $\delta p = (\frac{\partial p}{\partial n})\delta n +
(\frac{\partial p}{\partial\rho})\delta\rho$. 
 These hyperbolizations are all causal
provided
\begin{equation}
\frac{\partial p}{\partial\rho}+\frac{n}{\rho+p}\frac{\partial p}{\partial n
}\leq 1
\label{A.17}\end{equation}
(i.e., physically, provided the sound-speed does not exceed
light-speed).  If (\ref{A.17}) fails, then none are causal.  There are no constraints
(as follows, e.g., from existence of a hyperbolization and
equality of the dimension of the space of fields and that of equations.)

The allowed interactions (e.g., electromagnetic, contact-force, etc.)
are given by
\begin{equation}
j_A=(j_a,\; \;  j,\; \; 0).
\label{A.18}\end{equation}
That is, arbitrary sources are allowed in the equation of
stress-energy conservation, but none is allowed in the equation of
particle-number conservation. 
\\

\noindent{\bf Gravitation}

The fields consist of a symmetric, Lorentz-signature metric,
$g_{ab}$, together with a (torsion-free) derivative operator,
$\nabla_a$, on $M$.  The equations are
\begin{equation}
\nabla_ag_{bc} =0,
\label{A.19}\end{equation}
\begin{equation}
R_{ab(c}{^m}g_{d)m}=0,
\label{A.20}\end{equation}
\begin{equation}
R_{m(ab)}{^m}=0.
\label{A.21}\end{equation}
Here, $R_{abc}{^d}$, with symmetries $R_{abc}{^d} =
R_{[ab]c}{^d}$ and $R_{[abc]}{^d} =0$, is the curvature
tensor (i.e., the ``derivative of the derivative operator"),
defined by the condition that 
\begin{equation}
\nabla_{[a}\nabla_{b]}\xi_c=\frac{1}{2}R_{abc}{^d}\xi_d,
\label{A.22}\end{equation}
for every covector field $\xi_c$ on $M$.

Thus, the fibres of the bundle $b$ are fifty-dimensional,
a typical vertical vector being given by $\delta\phi^{\alpha'}
=(\delta g_{ab},\delta\Gamma^a{_{bc}})$, where
$\delta g_{ab} = \delta g_{(ab)}$ (ten dimensions), and
$\delta\Gamma^a{_{bc}}=\delta\Gamma^a{_{(bc)}}$ (forty
dimensions).  The latter represents a first-order
change in the derivative operator, whose effect is to replace $\nabla_a
\xi_b$ by $\nabla_a \xi_b + \delta\Gamma^m{_{ab}}\xi_m$. 
A typical vector in the space of equations is
$\sigma^A=(s^{abc}, s^{abcd}, s^{ab})$ (the respective coefficients
of (\ref{A.19})-(\ref{A.21})), where
$s^{abc}=s^{a(bc)}$, $s^{abcd}=s^{[ab](cd)}$, and
$s^{ab}=s^{(ab)}$.  Thus, the equation space has dimension
one hundred ten ($= 40+60+10$).

This system consists of two separate physical systems (in the
sense of Sect. 6).  For the first, the field is the metric
$g_{ab}$, and the equation (\ref{A.19}), with no background
field.  Its most general hyperbolization at a point is given by
\begin{equation}
h^A{_{\alpha'}}\delta\phi^{\alpha'}=x^{abcmn}
\delta g_{mn},
\label{A.23}\end{equation}
where $x^{abcmn}$ has symmetries $x^{abcmn}=x^{a(mn)(bc)}$,
and is such that $n_ax^{abcmn}$ is positive-definite in the
index pairs ``$b,c$" and ``$m,n$", for some $n_a$.  The general constraint
 for this physical system at a point is given by
\begin{equation}
c^{An}=x^{nabc},
\label{A.24}\end{equation}
where $x^{nabc}=x^{[na](bc)}$.    
Thus, the vector space of constraints has dimension sixty, while
the space of vectors of the form $c^{An}n_n$, for fixed
$n_n$, has dimension thirty.  These constraints are complete
and, by virtue of Eqn. (\ref{A.20}), integrable. For the other physical system, the field is
the derivative operator, the equations are (\ref{A.20}), (\ref{A.21}),
and the background field is the space-time metric $g_{ab}$.
This physical system has no hyperbolization, because of
the diffeomorphism freedom.  (But it does have a ``hyperbolization
up to diffeomorphisms".  See Eqn. (\ref{5.10})). The most general
constraint for this system is given by
\begin{equation}
c^{An}=(x^{nabcd}+2x^{[a}g^{b](c}g^{d)n}-
g^{cd}x^{[a}g^{b]n},\; \; \; x^{(a}g^{b)n} 
-\frac{1}{2} g^{ab}x^n),
\label{A.25}\end{equation}
where $x^{nabcd}=x^{[nab](cd)}$. 
Thus, the vector space of constraints has dimension
fifty-four, while the space of vectors of the form $c^{An}n_n$,
for fixed $n_n$, has dimension thirty-four.  These constraints are
integrable, but not complete.  (This feature, again, is
related to diffeomorphism-invariance.  See Sect. 5.)  

The most general allowed interaction, apparently, is
\begin{equation}
\delta j_A=(0,\; \; 0,\; \; T_{ab}-\frac{1}{2}T^m{_m}g_{ab}),
\label{A.26}\end{equation}
where $T_{ab} = T_{(ab)}$ (the stress-energy of matter).  In order
that this interaction preserve integrability of the
constraints, we must have conservation, $\nabla^bT_{ab}=0$.
This is a very severe restriction on the fields
contributing to $T_{ab}$, and the interactions between
those fields.   
\\

\noindent{\bf Spin-s Systems}

The field is a totally symmetric, $2s$-rank spinor,
$\psi^{A\cdots D}=\psi^{(A\cdots D)}$, on $M$, with background metric
$g_{ab}$.  Here, $s = \frac{1}{2}, 1, \frac{3}{2}, 2,\cdots$.  The equation is
\begin{equation}
\nabla^{A'}{_A}\psi^{AB\cdots D}=0.
\label{A.29}\end{equation}

Thus, the fibres of the bundle $b$ are ($4s+2$)-dimensional.
(This, and all subsequent, reference to dimension means
{\em real} dimension.) 
A typical vertical vector is given by $\delta\phi^{\alpha'}
=\delta\psi^{A\cdots D} = \delta\psi^{(A\cdots D)}$.  A typical vector
in the space of equations is given\footnote{The notation is a
little awkward here.  The index on the left is in the
space of equations; those on the right, in spinor space.} by $\sigma^{A} 
=s_{A'}{_{B\cdots D}}=s_{A'}{_{(B\cdots D)}}$.  Thus, the equation-space
is $8s$-dimensional.  

The most general hyperbolization at a point is given by
\begin{equation}
h^A{_{\alpha'}}\delta\phi^{\alpha'}=t_{B'\cdots D'B\cdots D}
\delta\bar{\psi} 
_{A'}{^{B'\cdots D'}},
\label{A.30}\end{equation}
where $t_{B'\cdots D'B\cdots D}=\bar{t}_{(B'\cdots D')
(B\cdots D)}$, a Hermitian quadratic form on symmetric,
($2s-1$)-rank spinors, is positive-definite. 
The
general constraint is given by
\begin{equation}
c^{An}=\epsilon^{N'A'}\epsilon^{N(B}x^{C\cdots D)},
\label{A.31}\end{equation}
where $x^{C\cdots D}$ is an arbitrary symmetric, ($2s-2$)-rank spinor.
Thus, the constraints form a ($4s-2$)-dimensional vector
space, while the space of vectors of the form $c^{An}n_n$, 
for fixed $n_n$, also has dimension ($4s-2$).  Note that, for $s=1/2$, there are no constraints. 
These constraints are always complete, but they are 
integrable if and only if either $s\leq 1$, or the metric
$g_{ab}$ is conformally flat.  (This is the famous ``inconsistency of
the higher-spin equations".)  Except for the cases $s=1$ (electromagnetism),
and $s=2$ (linearized gravity), it isn't clear what are the allowed interactions.  For charged spin-$s$ fields, see later in this Appendix.
\\

\noindent{\bf Elastic Solid}

The fields consist of a scalar field $\rho$, a unit
timelike vector field $u^a$, and a symmetric tensor field $h_{ab}$
of signature $(0,+,+,+)$ satisfying $h_{ab}u^b=0$, with background metric
$g_{ab}$.  The equations are
\begin{equation}
u^m\nabla_mh_{ab}+2(\nabla_{(a}u^m)h_{b)m}=0,
\label{A.32}\end{equation}
\begin{equation}
\rho u^m\nabla_mu^a+q^a{_b}\nabla_m\tau^{bm}=0,
\label{A.33}\end{equation} 
\begin{equation}
u^m\nabla_m\rho+\rho\nabla_mu^m+\tau^{mn}\nabla_mu_n=0,
\label{A.34}\end{equation}
where we have set $q_{ab}=g_{ab}+u_au_b$, the ``spatial metric".
Here, $\tau^{ab}$ is some fixed algebraic function
of $h_{ab}$, $u^a$, and $g_{ab}$, satisfying $\tau^{ab}u_b=0$ and
$\tau^{ab}=\tau^{(ab)}$. 
The physical meaning of these equations is the following.  The
field $\rho$ is the mass density, and $u^a$ the material
four-velocity.  The field $h_{ab}$ represents a sort of 
``natural spatial geometry" for the material.  Thus, we
interpret the combination $h_{ab}-q_{ab}$ (natural geometry
minus actual geometry) as the strain on the material;
and Eqn. (\ref{A.32}) (which is just ${\cal L}_u h_{ab}=0$) as
requiring that the
material carry along with it its natural geometry.
The field $\tau^{ab}$ represents the stress of the material.
Thus, we interpret $\tau^{ab}(h_{ab})$ as the stress-strain
relation\footnote{Note that the stress-strain relation
need not be linear.  It would perhaps be natural to require
that $\tau^{ab}$ vanish when $h_{ab}=q_{ab}$, but that
requirement is not needed for what follows.}, and
the combination  $\rho u^au^b+\tau^{ab}$
as the material stress-energy.  Eqns. (\ref{A.33}),
(\ref{A.34}) are precisely conservation of this
stress-energy.

This system admits a hyperbolization if and only if $\rho>0$;
and in addition the tensor
$\tau^{abcd} = \partial\tau^{ab}/\partial h_{bc}$
is symmetric under interchange of the index pairs
$a$,$b$ and $c$,$d$ and positive-definite
in these pairs. 
In this case, the most general hyperbolization is
given by
\begin{eqnarray}
h^A{_{\alpha'}}\delta\phi^{\alpha'}&=&x[2\delta\rho
-\tilde{h}^{cm}(\rho q^d{_m}+\tau^d{_m})\delta h_{cd}]
(-\tilde {h}^{an}(\rho q^b{_n}+\tau^b{_n}),\; 0,\; 2)\nonumber\\ 
&&+ y(\tau^{abcd}
\delta h_{cd},\; \; 2h^{am}\delta u_m,\; \; 0),  
\label{A.35}
\end{eqnarray}
where $x$ and $y$ are numbers with $xy>0$, and $\tilde{h}{^{ab}}
=\tilde{h}^{(ab)}$
is defined by $u_b\tilde{h}^{ab}=0$ and $\tilde{h}^{am}   
h_{bm} = q^a{_b}$.  These hyperbolizations are all causal if
and only if 
\begin{equation}
h_{mn}\tau^{ambn}
\leq\frac{1}{2}\rho q^{ab},
\label{A.36}\end{equation}
which means, physically, that no acoustic wave-speed
exceed the speed of light. 
This system has no constraints.  If we generalize this system
to allow the stress $\tau^{ab}$ to depend, not only on 
the natural geometry $h_{ab}$, but also on the mass density $\rho$, then the
Eqns. (\ref{A.32})-(\ref{A.34}) never admit a hyperbolization.
It seems peculiar that such a physically benign generalization
would preclude a hyperbolization.  What is going on here?

Presumably, the most general allowed interaction is given 
by
\begin{equation}
j_A=(0,\; \;j_a,\; \; j).
\label{A.37}\end{equation}
That is, interactions are allowed that exchange energy-momentum
with the environment, but not that modify Lie
transport of the natural geometry.
\\

\noindent{\bf Special Relativity}

The fields consist of a symmetric, Lorentz-signature metric
$g_{ab}$, together with a derivative operator
$\nabla_a$, on $M$.  The equations are
\begin{equation}
\nabla_ag_{bc}=0,
\label{A.38}\end{equation}
\begin{equation}
R_{abc}{^d}=0,
\label{A.39}\end{equation}
where $R_{abc}{^d}$ is the curvature tensor, given by (\ref{A.22}).

The bundle space $b$ is this case is identical to that for 
general relativity, and so in particular the
fibres have dimension fifty.  But now the equations are different.  A
typical vector in the space of equations is
$\sigma^A=(s^{abc},s^{abc}{_d})$ (respective coefficients of
(\ref{A.38}), (\ref{A.39})), where $s^{abc}=s^{a(bc)}$,
$s^{abc}{_d}=s^{[ab]c}{_d}$, and $s^{[abc]}{_d}=0$.  Thus,
the equation-space has dimension one hundred twenty
($=40+80$).

This system consists of two separate physical systems (in the 
sense of Sect. 6).  The first, with field the metric, is
identical to the similar system for general relativity.  Thus,
the hyperbolizations are given by (\ref{A.23}), the
constraints by (\ref{A.24}).  For the second system, the
field is the derivative operator, and the equation 
(\ref{A.39}).  This system has no hyperbolizations,
because of the diffeomorphism freedom.  The most general
constraint for this system is given by
\begin{equation}
c^{An}=x^{nabc}{_d},
\label{A.40}\end{equation}
with $x^{nabc}{_d}=x^{[nab]c}{_d}$ and $x^{[nabc]}{_d}=0$.
Thus, the vector space of constraints has dimension sixty, while
the space of vectors of the form $c^{An}n_n$, for fixed $n_n$,
has dimension forty-eight\footnote{Here is a mystery.  For
this system, the left side of Eqn. (\ref{4.4}) exceeds the
right side by eight.  But we might have expected, from the fact
that Eqns. (\ref{A.38}), (\ref{A.39}) have an initial-value
formulation up to diffeomorphisms, an excess of four.  What is
the explanation for this discrepency?} The constraints are
integrable, but not complete. 

Apparently, no interactions whatever are permitted in
Eqns. (\ref{A.38}), (\ref{A.39}).  Note that passage to
general relativity is not ``turning on an interaction", because
this is a  change in the dynamical part of Eqn. (\ref{A.39}).
\\

\noindent{\bf Dust}

The fields consist of a scalar field, $\rho$, and a unit
timelike vector field, $u^a$, on $M$, with
background metric $g_{ab}$.  The equations are
\begin{equation}
u^m\nabla_mu^a=0,
\label{A.41}\end{equation}
\begin{equation}
u^m\nabla_m\rho+\rho\nabla_mu^m=0.
\label{A.42}\end{equation}

Thus, the fibres of the bundle $b$ are four-dimensional, a
typical vertical vector being given by $\delta\phi^{\alpha'}
=(\delta u^a,\delta\rho)$, with $u_a\delta u^a=0$.  A typical 
vector in the space of equations is $\sigma^A=(s_a,s)$
(respective coefficients of (\ref{A.41}), (\ref{A.42})),
with $s_au^a=0$.  The equation-space is four-dimensional.

Remarkably enough, this system admits no hyperbolization.  To
see this, note that the most general candidate for a hyperbolization
at a point is
\begin{equation}
h^A{_{\alpha'}}\delta\phi^{\alpha'}=(x_{ab}\delta u^b
+x_a\delta\rho,\; \; y_a\delta u^a+y\delta\rho),
\label{A.43}\end{equation}
for some tensors $x_{ab}$, $x_a$, $y_a$, and $y$.  Combining this
with the $k_A{^m}{_{\alpha'}}$ from Eqns. (\ref{A.41}),
(\ref{A.42}), we obtain
\begin{eqnarray}
&&h^A{_{\alpha'}}\delta\phi^{\alpha'}k_A{^m}{_{\beta'}}
\delta\hat{\phi}^{\beta'}=u^m[x_{ab}\delta
\hat{u}^a\delta u^b +x_a\delta\hat{u}^a\delta\rho
+y_a\delta\hat{\rho}\delta u^a+y\delta\hat{\rho} 
\delta\rho]\nonumber\\
&&\ \ \ \ \ \ \ + \rho\delta\hat{u}^m
[y_a\delta u^a+y\delta\rho].
\label{A.44}\end{eqnarray}
We see that this is symmetric in $\delta\phi^{\alpha'}$ and
$\delta\hat{\phi}^{\beta'}$ if and only if $y$, $y_a$,
and $x_a$ all vanish, and $x_{ab}$ is symmetric. But these
conditions preclude positive-definiteness of (\ref{A.44}).
This lack of a hyperbolization is discussed briefly
at the end of Sect. 3.  There are no constraints.

Any $\delta j_A$ is, apparently, allowed as an interaction.
\\

\noindent{\bf Charged Fields}

Fix on $M$ a smooth, antisymmetric tensor field $F_{ab}$ having
vanishing curl ($\nabla_{[a}F_{bc]}=0$).  We assume that $M$ is simply
connected, and that $F$ has been so chosen that its integral
over every compact 2-surface in $M$ vanishes\footnote
{Vanishing of the 2-surface integrals of $F$ means
physically that all wormholes manifest zero net magnetic 
charge.  In fact, we could allow here integral multiples
of a certain fundamental magnetic charge, but,
for simplicity, do not.  The assumption of simple-connectedness
of $M$ avoids our having to consider Aharonov-Bohm effects.
Again, we could relax this assumption, at the cost of somewhat
complicating the discussion below.}.

We wish to introduce the notion of tensors on $M$ of charge $e$
(some fixed number) and index type $t^{a\cdots}{_{b\cdots}}$
(some fixed arrangement of contravariant and covariant $M$-indices).
To this end, fix a reference point $x_o$ of $M$, and consider any 
point $x$ of $M$.  Consider the collection of all pairs,
($A_a,t^{a\cdots}{_{b\cdots}}$), where $A_a$ is a
vector field on $M$ with $\nabla_{[a}A_{b]}=F_{ab}$ (existence
of which is guaranteed by the assumptions above), and 
$t^{a\cdots}{_{b\cdots}}$ is a complex tensor at $x$.
Two such pairs, ($A_a,t^{a\cdots}{_{b\cdots}}$), and
$(A'_a,t'^{a\cdots}{_{b\cdots}}$), are taken as equivalent
if
\begin{equation}
t'^{a\cdots}{_{b\cdots}}=\exp[ie\int(A'_m-A_m) ds^m]
\; \; \; t^{a\cdots}{_{b\cdots}},
\label{A.45}\end{equation}
where the integral on the right is over any curve from
$x_o$ to $x$.  (Independence of the choice of curve is
guaranteed by the conditions, above, on $A_a$ and $A'_a$.)
This is an equivalence relation.  The equivalence classes,
called {\em charge-e tensors} at $x$, form a vector space
of (complex) dimension $4^s$, where $s$ is the total number
of indices of $t^{a\cdots}{_{b\cdots}}$.  
A charge-$e$ tensor field is an assignment, to each point
$x\in M$, of a charge-$e$ tensor at $x$.  Thus, a charge-$e$ tensor
field can be represented by a pair ($A_a,t^{a\cdots}{_{b\cdots}}$)
of fields on $M$, where pairs ($A_a,t^{a\cdots}{_{b\cdots}}$)
and ($A'_a,t'^{a\cdots}{_{b\cdots}}$) are identified provided
Eqn. (\ref{A.45}) holds for all $x$. 

The discussion above was predicated on the choice of a fixed
reference point, $x_o\in M$.  Let us now change to a new
reference point, $\tilde{x}_o\in M$. Fix a smooth curve
$\gamma$ from $x_o$ to $\tilde{x}_o$.  Then we identify a
charge-$e$ tensor at $x$, defined via reference point
$x_o$, with charge-$e$ tensor at $x$, defined via reference
point $\tilde{x}_o$, provided these have respective
representatives, ($A_a,t^{a\cdots}{_{b\cdots}}$) and
($A_a,\tilde{t}^{a\cdots}{_{b\cdots}}$), with
$\tilde{t}^{a\cdots}{_{b\cdots}}=\exp[ie\int_{\gamma}A_m ds^m]\; t^{a\cdots}{_{b\cdots}}$.  Note that
this is independent of representative.  But it does depend
on choice of the curve $\gamma$:  A change in $\gamma$
changes this identification by an overall phase (the same
for each point $x$).  Thus, charged tensor fields (independent
of reference point) make sense only up to an overall
constant phase.

Now let there be given on $M$ a space-time metric $g_{ab}$, with
corresponding derivative operator $\nabla_a$.  We extend the action
of this derivative operator to charged fields as follows.
Given a charge-$e$ tensor field, choose any representative
($A_a,t^{a\cdots}{_{b\cdots}}$), of it, and let its
derivative be the charge-$e$ tensor field with representative
($A_a,(\nabla_c-ieA_c)t^{a\cdots}{_{b\cdots}}$), noting that
this is independent of the original choice of representative.

The charged Klein-Gordon system consists of charge-$e$ fields
$\psi$ and $\psi_a$, with background fields the metric
$g_{ab}$ and electromagnetic field $F_{ab}$.  The
equations are the same as (\ref{A.6})-(\ref{A.8}), where, 
of course, $\nabla_a$ now is the derivative operator on
charged fields, except
that there appears on the right of Eqn. (\ref{A.7}) the term
$ie\psi F_{ab}$.  (Failure to include this term
would destroy integrability of the constraints.)  Similarly,
the charged spin-s system consists of a charge-$e$ spinor
field, $\psi^{A\cdots D}$, satisfying Eqn. (\ref{A.29}).

In order to fit these systems into the present framework, we
must fix a reference point $x_o$, for otherwise the freedom
to multiply fields by an overall constant phase makes it
impossible to find a bundle to house the charged fields.  The discussion of
hyperbolizations and constraints then goes through in a manner
identical to that of the uncharged case, with one exception.
The constraints for the equation for charged spin-$s$
fields, for $s\geq 1$ and $F_{ab}\neq 0$, fail to be integrable. 
One further complication arises.  There is
no natural way to lift diffeomorphisms from $M$ to the bundle
spaces $b$ associated with any of these charged fields (although there
does, of course, exist a lift ``up to overall
constant phase"). The reason for this is our having fixed a reference
point, $x_o$, in order to introduce $b$.
\\

\noindent{\bf Kinetic Theory}

Let $M$, $g_{ab}$ be a time-oriented space-time.  Fix a nonnegative number
$m$, 
and denote by $\Gamma$ the seven-dimensional manifold of
all pairs ($x$, $p_a$), where $x\in M$, and $p_a$ is a
future-directed vector at $x$ with
$p_ap^a=-m^2$.  Thus, $\Gamma$ is a fibre bundle over $M$.
Denote by $\Gamma_x$ the fibre over $x\in M$, so each
$\Gamma_x$ is a three-manifold.  The field for
kinetic theory is a nonnegative function $f$ on the
manifold $\Gamma$; and the equation is
\begin{equation}
p^a\nabla_af={\cal C}(f).
\label{A.46}\end{equation}
Here, $\cal C$ is, for each $x\in M$, a mapping from nonnegative functions
$h$ on $\Gamma_x$ to functions on $\Gamma_x$, satisfying
\begin{equation}
\int{\cal C}(h)p^a=0,\; \; \; \int{\cal C}(h)p^ap^b=0,
\label{A.47}\end{equation}
for every $h$, where the integrals are over $\Gamma_x$
using the natural volume element on this mass shell.  Eqn
(\ref{A.46}) is to hold for each ($x$, $p_a$)$\in\Gamma$,
with $\cal C$ evaluated on the restriction of $f$
to $\Gamma_x$.  The physical interpretation of these
equations is the following.  The nonnegative function
$f$ is the distribution function (of particle
position-momentum) for mass-$m$ particles, and Eqn. (\ref{A.46}) is Boltzmann's
equation.  The $\cal C$ in (\ref{A.46}) is the collision
function; and (\ref{A.47}) is local conservation, in
collisions, of particle-number and energy-momentum.

This system does not, strictly speaking, fall within the
famework of Sect. 2, for the space of allowed ``field values"
at each point of $M$ (nonnegative functions on $\Gamma_x$)
is infinite-dimensional.  But, if we agree to ignore
this one defect, there results a nice example of our framework.
  A typical
vertical vector is given by $\delta\phi^{\alpha'}=\delta f$, 
a function on $\Gamma_x$.  A typical vector in the space of
equations is $\sigma^A=s$, a function on $\Gamma_x$.
The most general hyperbolization of (\ref{A.46}) is given by
\begin{equation}
h^A{_{\alpha'}}\delta\phi^{\alpha'}=\mu\; \delta f,
\label{A.48}\end{equation}
where $\mu$ is any positive function on $\Gamma_x$.  There are no
constraints.  Presumably, there is allowed as an interaction
in (\ref{A.46}) any function $\delta j$ on $\Gamma$
such that, for every $x\in M$, $\int(\delta j)p^a=0$,
where this integral is over $\Gamma_x$.  That is,
interactions may not violate local particle-number
conservation, but are otherwise arbitrary. 
\\

\noindent{\bf Lagrangian Systems}

Fix a fibre bundle $\hat{b}\stackrel{\hat{\pi}}{\rightarrow}M$.
By a (first-order) {\em Lagrangian} on the bundle
$\hat{b}$, we mean a smooth function $L$ as follows:  $L$
is a function on pairs, ($\kappa$, $\zeta_a{^{\alpha}}$),
where $\kappa$ is a point of the bundle space $\hat{b}$,
and $\zeta_a{^{\alpha}}$ is a tensor at $\kappa$ satisfying
$\zeta_a{^{\alpha}}(\nabla\pi)_{\alpha}{^b}=\delta_a{^b}$;
and, for each such pair, $L(\kappa,\zeta_a{^{\alpha}})$ is
a density\footnote{That is, an antisymmetric, fourth-rank
$M$-tensor, whose indices we shall suppress.} in $M$ at
the point $\pi(\kappa)$.  Such a Lagrangian gives rise
to a system of partial differential equations on
cross-sections $\hat{\phi}$ of the bundle $\hat{b}$.
In order to write these equations explicitly, it is convenient
to introduce a derivative operator $\nabla_{\alpha}$ on
mixed fields on $\hat{b}$, such that the derivative
of every vertical vector field is vertical, and
$\nabla_{\alpha}(\nabla\pi)_{\beta}{^a}=0$.  Then,
e.g., the operator ``derivative along the cross-section
$\hat{\phi}$" is given by $(\nabla\hat{\phi})_a{^{\alpha}}
\nabla_{\alpha}$.
(See the discussion just preceeding Eqn. (\ref{4.5}).)
Written in terms of this operator, Lagrange's equation
becomes\footnote{Note that the Greek index of 
``$\partial/\partial\zeta_m{^{\alpha'}}$" acquires a prime,
as a consequence of the fact that $\zeta_a{^{\alpha}}
(\nabla\pi)_{\alpha}{^b}=\delta_a{^b}$.} 
\begin{equation}
\frac{\partial^2L}{\partial\zeta_m{^{\alpha'}}\partial
\zeta_n{^{\beta'}}}(\nabla\hat{\phi})_m{^{\mu}}
\nabla_{\mu}((\nabla\hat{\phi})_n{^{\beta}})
+(\nabla\hat{\phi})_m{^{\beta}}\nabla_{\beta}
(\frac{\partial L}{\partial\zeta_m{^{\alpha'}}}) 
-\nabla_{\alpha'}L=0.
\label{A.49}\end{equation}
The coefficients in (\ref{A.49}) are to be evaluated at 
$\zeta_a{^{\alpha}} = (\nabla\hat{\phi})_a{^{\alpha}}$.
This equation is of course independent of the
choice of derivative operator.  It is also unchanged under adding
to $L$ any function of the form $(\nabla_{\alpha}v^a)\zeta_a{^{\alpha}}$
(a ``total divergence"), where $v^a$ is any $M$-density field
on $\hat{b}$.  Note that the spaces of equations and unknowns
for Eqn. (\ref{A.49}) have the same dimension,
namely, that of the fibres of the bundle $\hat{b}$.

In order to cast Eqn. (\ref{A.49}) into first-order form,
we introduce auxiliary field $\zeta_a{^{\alpha}}$, subject
to $\zeta_a{^{\alpha}}(\nabla\pi)_{\alpha}{^b}
=\delta_a{^b}$; and we supplement (\ref{A.49}) with the
additional equations
\begin{equation}
(\nabla\hat{\phi})_a{^{\alpha}}=\zeta_a{^{\alpha}},
\label{A.50}\end{equation}
\begin{equation}
\zeta_{[b}{^{\beta}}\nabla_{|\beta |}\zeta_{a]}{^{\alpha}}
=0.
\label{A.51}\end{equation}
The system (\ref{A.50}), (\ref{A.51}), (\ref{A.49}) is closely analogous
to the Klein-Gordon system, (\ref{A.6})-(\ref{A.8}).
 
The fibres of the bundle $b$ appropriate to this system have
dimension $5n$, where $n$ is the dimension of the fibres of
$\hat{b}$.  A typical vertical vector in $b$ is given by
$\delta\phi^{\alpha'}=(\delta\hat{\phi}^{\alpha'},\delta
\zeta_a{^{\alpha'}})$.  Here, $\delta\hat{\phi}^{\alpha'}$, 
a vertical vector in $\hat{b}$, represents an infinitesimal
change in the value of the cross-section $\hat{\phi}$ of
$\hat{b}$, while $\delta\zeta_a{^{\alpha'}}$ represents
an infinitesimal change in the tensor $\zeta_a{^{\alpha}}$.
A typical vector in the space of equations is
$\sigma^A=(s^a{_{\alpha'}}, s^{ab}{_{\alpha'}}, s^{\alpha'})$,
with\footnote{Note that the difference between the
two sides of (\ref{A.50}), as well as the left side of (\ref{A.51}),
is automatically vertical.  Thus, a prime must be attached to the Greek
subscripts of $s^a{_{\alpha'}}$ and $s^{ab}{_{\alpha'}}$.}  $s^{ab}{_{\alpha'}}= s^{[ab]}{_{\alpha'}}$ (respective
coefficients of (\ref{A.50}), (\ref{A.51}), (\ref{A.49}))
Thus, the equation-space has dimension $11n$ ($=4n+6n+n$).

Set
\begin{equation}
S^{ab}{_{\alpha'\beta'}}=
\frac{\partial^2L}{\partial\zeta_{(a}{^{\alpha'}}
\partial\zeta_{b)}{^{\beta'}}}, 
\label{A.52}\end{equation}
the coefficient of the first term in (\ref{A.49}).  The
system (\ref{A.49})-(\ref{A.51}) admits a hyperbolization at a point
if and only if the tensor $S^{ab}{_{\alpha'\beta'}}v^{\alpha'}
v^{\beta'}$ is Lorentz-signature for every nonzero
$v^{\alpha'}$, and, in addition, there exists a tangent vector
$t^a$ and a covector $n_a$ that are both timelike for every one
of these Lorentz metrics.  Note that this is a rather
severe condition on $S^{ab}{_{\alpha'\beta'}}$.  When it is
satisfied, the 
most general hyperbolization is given by
\begin{equation}
h^A{_{\alpha'}}\delta\phi^{\alpha'}=
(-w^a{_{\alpha'\beta'}},\; -t^{[a}S^{b]m}{_{\alpha'\beta'}}
\delta\zeta_m{^{\beta'}},\; \frac{1}{2}t^m\delta\zeta_m{^{\alpha'}}), 
\label{A.53}\end{equation}
where $t^a$ is such a common timelike vector, and $w^a{_{\alpha'\beta'}}
=w^a{_{(\alpha'\beta')}}$ has the property that 
$(t^bn_b)n_aw^a{_{\alpha'\beta'}}$ is positive-definite
for some such common timelike covector $n_a$.  (Such a
$w^a{_{\alpha'\beta'}}$ always exists, e.g., that 
given by $t^aG_{\alpha'\beta'}$ with $G_{\alpha'\beta'}$
positive-definite.)  Given a space-time metric $g_{ab}$, this
system is causal if and only if every $n_a$ lying in
one half of the light cone of $g_{ab}$ will serve as
a ``common timelike covector" above. 
 Eqn. (\ref{A.53}) should be compared
with Eqn. (\ref{A.9}), giving the Klein-Gordon hyperbolizations.  
The most general constraint for the system (\ref{A.49})-(\ref{A.51})
is given by
\begin{equation}
c^{An}=(x^{na}{_{\alpha'}},\;  y^{nab}{_{\alpha'}},\;  0),
\label{A.54}\end{equation}
where $x^{na}{_{\alpha'}}=x^{[na]}{_{\alpha'}}$ and
$y^{nab}{_{\alpha'}}=y^{[nab]}{_{\alpha'}}$.  Thus,
the constraints form a vector space of dimension
$10n$, while the space of vectors of the form $c^{An}n_n$,
for fixed $n_n$, has dimension $6n$.  Eqn. (\ref{A.54})
should be compared with Eqn. (\ref{A.10}), giving the Klein-Gordon
constraints.  These constraints
are complete and integrable.

This system only allows those interactions that preserve its
Lagrangian character, i.e., that result from some change
in the Lagrangian.  This change must be so chosen to leave
invariant
the tensor $S^{ab}{_{\alpha'\beta'}}$ of Eqn. (\ref{A.52})
(this being the coefficient of the dynamical term
in (\ref{A.49})).  The most general such change 
is given by $\delta L=W^m{_{\alpha}}\zeta_m{^{\alpha}}
+W$, where $W^m{_{\alpha}}$ and $W$ are arbitrary
smooth fields on the bundle space $\hat{b}$.  That is,
these $W$'s are allowed to depend on the $\kappa$-variables,
but not on the $\zeta_a{^{\alpha}}$.  This change in the Lagrangian $L$
results in a term 
\begin{equation}
\delta j_A=(0,\; 0,\; 2(\nabla_{[\alpha'}W_{\beta]}{^m})
\zeta_m{^{\beta}}-\nabla_{\alpha'}W)
\label{A.55}\end{equation}
in (\ref{2.1}).   
Thus, the allowed interactions on a Lagrangian system are
very special indeed.  In particular, the $\delta j_A$ 
can be at most linear in the field $\zeta_a{^{\alpha}}$.

One example to which the discussion above can be applied
is the Klein-Gordon system, with Lagrangian
$L=\frac{1}{2}(\nabla_a\psi)(\nabla^a\psi)$.  But there
exists an alternative Lagrangian formulation for this
system, starting from $L=-\frac{1}{2}\psi_a\psi^a+
\psi^a\nabla_a\psi$.  This alternative Lagrangian involves
the full set of Klein-Gordon fields, $\psi$, $\psi_a$
(as opposed to just $\psi$), and yields equations that are
automatically first-order (as opposed to second-order).
It turns out that such an alternative Lagrangian
formulation is available quite generally.  We sketch below
how this comes about.

Recall that the points of the bundle $\hat{b}$ are denoted
$\kappa$, and that a Lagrangian on $\hat{b}$ consists of 
a certain function $L(\kappa,\; \zeta_a{^{\alpha}})$.
Fix such a Lagrangian.  Consider now the bundle $b$, whose
points are pairs ($\kappa$, $\zeta_a{^{\alpha}}$).  Let
us now introduce, on this new bundle $b$, the following
Lagrangian: 
\begin{equation}
\tilde{L}=(\frac{\partial L}{\partial\zeta_a{^{\alpha}}})
(\nabla\hat{\phi})_a{^{\alpha}}-L.
\label{A.56}\end{equation}
The Lagrange equations arising from (\ref{A.56}) are precisely
(\ref{A.49}) and (\ref{A.50}), except that in (\ref{A.49})
we must replace ``$(\nabla\hat{\phi})_a{^{\alpha}}$"
everywhere by ``$\zeta_a{^{\alpha}}$".  Thus, we obtain from
the Lagrangian (\ref{A.56}) a first-order system from the start.
But this system---despite the fact that its spaces of unknowns
and equations have the same dimension---admits no hyperbolization.
In fact, this system has constraints---essentially those given
by the $x^{na}{_{\alpha'}}$ in Eqn. (\ref{A.54}).  These
constraints are not integrable, but their integrability
conditions are (\ref{A.51}), which are linear in first
derivatives of the fields (as opposed to quadratic, which is
what happens in the general case, (\ref{4.5})).  So, we may
include these integrability conditions with the other
equations of our system. 
The result is Eqns. (\ref{A.49})-(\ref{A.51}),
a system that admits a hyperbolization and has complete, integrable
constraints.  That is, the result is a system with an initial-value formulation.
\\

\noindent{\bf Higher-Order Systems}

It is conceivable that some physical phenomena may be described by
higher-order systems of partial differential equations (e.g.,
arising from a Lagrangian of higher order).  We describe briefly
the conversion of such systems to first-order form, and their
resulting initial-value formulation.

Consider a quasilinear, $s^{th}$-order system of partial 
differential equations, which we may write in the form
\begin{equation}
k_A{^{m_1\cdots m_s}}{_{\alpha'}}\nabla_{m_1}\cdots\nabla_{m_s}
\phi^{\alpha'}+j_A=0,
\label{HO.1}\end{equation}
where $k_A{^{m_1\cdots m_s}}{_{\alpha'}}=k_A{^{(m_1\cdots m_s)}}
{_{\alpha'}}$ and $j_A$ are functions of point of $M$, the
field $\phi$, and, at most, its first ($s-1$) derivatives.
To achieve this form, we have introduced a connection in
the bundle of which $\phi$ is a cross-section.  (The coefficient
$k$ in (\ref{HO.1}), but not $j$, is independent of this 
choice.)  As an example, Lagrange's equation for a higher-order
Lagrangian takes the form (\ref{HO.1}), with the index
``$A$" replaced by ``$\beta'$", with 
$k_{\beta'}{^{m_1\cdots m_s}}{_{\alpha'}}$ symmetric in
$\beta',\alpha'$, and with s even.  Let us now, in order to achieve a 
first-order system, introduce auxiliary fields,
$\phi_a{^{\alpha'}}$, $\phi_{ab}{^{\alpha'}}$,$\cdots$,
$\phi_{a_1\cdots a_{s-1}}{^{\alpha'}}$, each symmetric in
its Latin indices, and their corresponding equations,
\begin{equation}
\nabla_{(a_1}\phi_{a_2\cdots a_i)}{^{\alpha'}}=
\phi_{a_1\cdots a_i}{^{\alpha'}},
\label{HO.2}\end{equation}
\begin{equation}
\nabla_{[a}\phi_{a_1]a_2\cdots a_i}{^{\alpha'}}=
\mu_{aa_1\cdots a_i}{^{\alpha'}}.
\label{HO.3}\end{equation}
Here, $i=1,\cdots,(s-1)$, and the $\mu$ on the right in
(\ref{HO.3}) is a certain function of the lower-order
$\phi$'s.  Our first-order system consists of Eqn. 
(\ref{HO.1}) with the derivative-term
replaced by ``$\nabla_{m_1}\phi_{m_2\cdots m_s}{^{\alpha'}}$",
and Eqns. (\ref{HO.2}), (\ref{HO.3}). 
The constraints for this system are always complete, and,
by virtue of the choice of $\mu$ in (\ref{HO.3}), integrable.

When does this system admit a hyperbolization?  No simple,
general criterion is known, but the following remarks will at
least suggest a possible line of attack.  First note that
the equations on $\phi_{a_1\cdots a_i}{^{\alpha'}}$ for
$i<(s-1)$ (namely, (\ref{HO.3}) for
$i<(s-1)$, and (\ref{HO.2})) always admit a hyperbolization (in a manner similar
to that of the Klein-Gordon system, (\ref{A.6}) and (\ref{A.7})).
What remains is the field $\phi_{a_1\cdots a_{s-1}}{^{\alpha'}}$,
and its equations, (\ref{HO.3}) (for $i=(s-1)$) and 
(\ref{HO.1}).  Let $t^A{_{\alpha'}}$ be any tensor such that
$t^A{_{\alpha'}}k_A{^{m_1\cdots m_s}}{_{\beta'}}$ is symmetric
in $\alpha'$,$\beta'$, and let $t^{a_1\cdots a_{s-1}}$ be
any totally symmetric tensor.  Consider the tensor given
by
\begin{eqnarray}
P_{\alpha'}{^{m_1\cdots m_{s-1}an_a\cdots n_{s-1}}}{_{\beta'}}
&=&\frac{1}{2}t^A{_{\alpha'}}t^{\{ m_1\cdots m_{s-1}}
k_A{^{n_1\cdots n_{s-1}\} a}}{_{\beta'}}\nonumber\\ 
&&-\frac{1}{2} t^A{_{\alpha'}}t^{a\{ m_1\cdots m_{s-2}}
k_A{^{m_{s-1}n_1\cdots n_{s-1}\} }}{_{\beta'}},  
\label{HO.4}\end{eqnarray}
where ``$\{ m_1\cdots m_{s-1}n_1\cdots n_{s-1}\} $"
means ``add all $2(s-1)$ terms that result from cyclic
permutations of these indices, and then
symmetrize the result over $m_1\cdots m_{s-1}$ and
over $n_1\cdots n_{s-1}$". 
Then, as is
easily checked directly, this $P$ has the
properties
\begin{equation}
 P_{\alpha'}
{^{m_1\cdots m_{s-1}an_1\cdots n_{s-1}}}{_{\beta'}}
=P_{(\alpha'}{^{(n_1\cdots n_{s-1})a(m_1\cdots m_{s-1})}}{_{\beta')}},
\label{HO.5}\end{equation}
\begin{equation}
P_{\alpha'}{^{m_1\cdots m_{s-1}(an_1\cdots n_{s-1})}}
{_{\beta'}}=t^{m_1\cdots m_{s-1}}t^A{_{\alpha'}}
k_A{^{n_1\cdots n_{s-1}a}}{_{\beta'}}. 
\label{HO.6}\end{equation}
 It follows from
Eqn. (\ref{HO.5}) that the differential operator
$P_{\alpha'}{^{m_1\cdots m_{s-1}an_1\cdots n_{s-1}}}{_{\beta'}} 
$ $\nabla_a\phi_{n_1\cdots n_{s-1}}{^{\beta'}}$ is automatically
symmetric; and, from Eqn. (\ref{HO.6}), that this differential
operator is a linear combination of the left sides of
(\ref{HO.3}) and (\ref{HO.1}).  Thus, we obtain in this way
a symmetrization of our first-order system.  
This
symmetrization is actually a hyperbolization provided we
can choose $t^A{_{\alpha'}}$ and $t^{a_1\cdots a_{s-1}}$
such that there exists at each point a covector
$n_a$ for which the quadratic form $n_aP_{\alpha'}
{^{m_1\cdots m_{s-1}an_1\cdots n_{s-1}}}{_{\beta'}}$ is
positive-definite (i.e., is positive when applied to any
nonzero $\delta\phi_{m_1\cdots m_{s-1}}{^{\alpha'}}$).

When can this positive-definiteness condition be achieved?  For $s=1$
(i.e., when (\ref{HO.1}) is already first-order), this condition
just repeats the definition of a hyperbolization of a 
first-order system.  The complete solution for $s=2$ is
given in the discussion of Lagrangian systems in this Appendix.
The cases $s>2$ are considerably more difficult.  Those
for odd and even $s$ appear to be rather different in
character.  Are there any simple, general conditions
on $k_A{^{m_1\cdots m_s}}{_{\alpha'}}$, for $s>2$,
that imply existence of $t^A{_{\alpha'}}$ and 
$t^{a_1\cdots a_{s-1}}$ yielding, by the construction above,
a hyperbolization?  Are there any other symmetrizations
of the first-order system (\ref{HO.1})-(\ref{HO.3})?

It is very likely (depending in part on how the derivative
in (\ref{HO.2}) is structured) that these higher-order
fields would, under infinitesimal diffeomorphisms, pick
up in (\ref{5.3}) higher derivatives of the generating
vector field $\xi^a$.  Thus, as discussed in Sect. 5,
such fields may usurp the diffeomorphism gauge group
from general relativity.

\section*{Appendix B---Existence and Uniqueness of\\
 Solutions of Symmetric, Hyperbolic Systems} 

Fix a smooth, four-dimensional manifold $M$, and a finite-dimensional
vector space $F$ (tensors over which will be denoted, respectively,
by Latin and Greek indices).  We are interested in $F$-valued
functions on $M$, $M\stackrel{\phi}{\rightarrow}F$.  Consider,
on such functions, the partial differential equation
\begin{equation}
k_{\alpha}{^m}{_{\beta}}(x,\phi^{\gamma}(x))\nabla_m\phi^{\beta}
+j_{\alpha}(x,\phi^{\gamma}(x))=0,
\label{B.1}\end{equation}
where $x\in M$, and $k_{\alpha}{^m}{_{\beta}}=k_{(\alpha}{^m}{_{\beta)}}$
and $j_{\alpha}$ are smooth functions on $M\times F$.
This is just Eqn. (\ref{2.1}), modified as follows: i) we have written the
bundle space as a product, and imposed a vector-space
structure on the fibres (which can always be done locally);
ii) we have chosen the gauge so that the last index of
$k$ is vertical, and then, since now all Greek indices are
vertical, have suppressed all primes; and iii) we have
multiplied Eqn. (\ref{2.1}) through by a suitable hyperbolization,
as in (\ref{3.3}). By {\em initial data} for (\ref{B.1}), we 
mean a three-dimensional submanifold $S$ of $M$, together
with a smooth mapping $S\stackrel{\phi_0}{\rightarrow}F$,
such that, for every point $x$ of $S$, the tensor
$n_m\; k_{\alpha}{^m}{_{\beta}}(x,\phi_0(x))$ is 
positive-definite, where $n_m$ is a normal to $S$ at $x$.
(This agrees with the definition of Sect. 3.)

The fundamental theorem\footnote{F. Johns, {\em Partial Differential
Equations}, Springer-Verlag (New York), l982; P.D. Lax, {\em Comm Pure and
Appl Math 8}, 615 (1955); K. O. Friedrichs, {\em Comm Pure and
Appl Math 7}, 345 (1954).} on existence and uniqueness of
solutions of symmetric, hyperbolic partial differential equations
may now be stated as follows.
\\

\noindent{\bf Theorem}.  Let $S\stackrel{\phi_0}{\rightarrow}F$
be initial data for (\ref{B.1}).  Then: 

a.  For some open neighborhood $U$ of $S$, there exists a smooth
solution, $U\stackrel{\phi}{\rightarrow}F$, of (\ref{B.1}), with
$\phi|_S=\phi_0$.

b.  For $U\stackrel{\phi}{\rightarrow}F$ and
$U'\stackrel{\phi'}{\rightarrow}F$ two such solutions
(so $\phi|_S=\phi'|_S=\phi_0$), there exists a neighborhood
$\hat{U}\subset U\cap U'$ of $S$ in which $\phi=\phi'$.

c.  Any $\hat{U}\subset U\cap U'$ will serve for part b
provided it can be covered by a smooth, one-parameter
family, $S_t$, of three-submanifolds of $U$ such that
i) one of the $S_t$ is $S$ itself, ii) each $S_t$
coincides with $S$ outside of a compact subset of $S$,
and iii) for every $t$, both $\phi$ and $\phi'$,
restricted to $S_t$, are initial data.
\\

Part a of this theorem is existence:  It guarantees a solution
of (\ref{B.1}), satisfying the given initial conditions, in
some neighborhood of the initial surface $S$.  Part b is
uniqueness:  It guarantees that two solutions, each
defined in some neighborhood of $S$, must coincide
in some common subneighborhood, $\hat{U}$.  Part c strengthens
part $b$ by guaranteeing a certain minimum ``size" for
$\hat{U}$:  Part b holds for any $\hat{U}$ that can be covered
by a one-parameter family, $S_t$, of surfaces  
 that result from deforming
$S$ within compact regions, provided each $S_t$
can serve as an initial-data surface\footnote{This
$\hat{U}$ lies within the domain of dependence of
$S$, suitably defined.  It is merely for convenience that we
characterize $\hat{U}$ using deformations of $S$, rather than
curves whose tangents are propagation directions, as defined
in Sect. 3.}.  Thus, given
$U\stackrel{\phi}{\rightarrow}F$ and
$U'\stackrel{\phi'}{\rightarrow}F$, we can generate
a $\hat{U}$ that works for part b by taking
compact deformations of $S$ within the intersection
$U\cap U'$, stopping as soon as positive-definiteness
of $n_m\; k_{\alpha}{^m}{_{\beta}}(x,\phi(x))$ or
$n_m\; k_{\alpha}{^m}{_{\beta}}(x,\phi'(x))$ fails.
It follows in particular from part c that the solution
at a point of this $\hat{U}$ depends only on the data in
a certain compact region of $S$. This observation is the 
basis of our discussion of ``signal-propagation directions"
in Sect. 3. We remark that there are also results that
strengthen part (a), by guaranteeing a certain minimum
``size" for its neighborhood $U$, but these are more
complicated and less useful than the strengthening of
part b, as given by part c above.   Of course,
solutions of Eqn. (\ref{B.1}) can---and often do---evolve
to become singular.

A sketch of the proof of the theorem follows.

Fix an alternating tensor, $\epsilon_{abcd}=\epsilon_{[abcd]}$,
on $M$ (to facilitate integration and to allow us to
take divergences of vector fields), and a positive-definite
metric $G_{\alpha\beta}$ on $F$ (to facilitate taking
norms).  We now derive an inequality, (\ref{B.3}), that we shall use
three times in what follows. 
We first note that, for
any fields ${\phi}^{\alpha}$,
${\phi}'^{\alpha}$, $\overline{k}_{\alpha}{^m}{_{\beta}}
= \overline{k}_{(\alpha}{^m}{_{\beta)}}$, $\overline{k}'
_{\alpha}{^m}{_{\beta}}=\overline{k}'_{(\alpha}{^m}
{_{\beta)}}$, $\overline{j}_{\alpha}$, and $\overline{j}'
_{\alpha}$ on $M$, we have   
\begin{eqnarray} 
&&\nabla_m[(\overline{k}_{\alpha}{^m}{_{\beta}}
+\overline{k}'_{\alpha}{^m}{_{\beta}})
(\phi^{\alpha}-\phi'^{\alpha})(\phi^{\beta}-\phi'^{\beta})]
\ \ \ \ \ \ \ \ \nonumber\\
&&\ =[\nabla_m(\overline{k}_{\alpha}{^m}{_{\beta}}
+\overline{k}'_{\alpha}{^m}{_{\beta}})]
(\phi^{\alpha}-\phi'^{\alpha})(\phi^{\beta}-\phi'^{\beta})\nonumber\\
&&\ \ \ -2(\phi^{\alpha}-\phi'^{\alpha})[\overline{k}_{\alpha}{^m}{_{\beta}}
-\overline{k}'_{\alpha}{^m}{_{\beta}}
]\nabla_m(\phi^{\beta}+\phi'^{\beta})\nonumber\\
&&\ \ \ -4(\phi^{\alpha}-\phi'^{\alpha})[\overline{j}_{\alpha}
-\overline{j}'_{\alpha}
]\nonumber\\
&&\ \ \ +4(\phi^{\alpha}-\phi'^{\alpha})[\overline{k}_{\alpha}{^m}{_{\beta}}
\nabla_m\phi^{\beta}+\overline{j}_{\alpha}
-\overline{k}'_{\alpha}{^m}{_{\beta}}
\nabla_m\phi'^{\beta}
-\overline{j}'_{\alpha}],
\label{B.2}\end{eqnarray}
where we have everywhere suppressed the variable $x$. 
 To prove Eqn. (\ref{B.2}), expand the left side, and note that all terms
cancel. Next, let $S_t$ ($t\in[0,t_o]$) be a smooth family of
three-submanifolds of $M$, each of which coincides with
$S=S_0$ outside a compact subset of $S$, and on each of which 
$n_m(\overline{k}_{\alpha}{^m}{_{\beta}}+\overline{k}'
_{\alpha}{^m}{_{\beta}})$ is positive-definite,
where $n_n$ is the normal to $S_t$ in the direction
of increasing $t$. 
Denote by $V$ the union of the $S_t$, so $V$ has a boundary
consisting of $S$ and $S_{t_o}$.  Now multiply
(\ref{B.2}) by exp($-2t/\tau$), where $\tau$ is
a positive number, and integrate over $V$.  Integrating the left
side by parts, the resulting volume integral involves the
expression $\frac{1}{\tau}(\nabla_m t)(\overline{k}_{\alpha}{^m}{_{\beta}}
+\overline{k}'_{\alpha}{^m}{_{\beta}})$.   
But, by construction, this tensor is positive-definite,
and so this volume integral will, provided $\tau$ is 
chosen sufficiently small, dominate the integral of the
first term on the right.  For the remaining three terms
on the right in Eqn. (\ref{B.2}), use the Schwarz inequality.
There results
\begin{eqnarray}
||(\phi-\phi')e^{-t/\tau}||
&\leq&\sqrt{\sigma\tau}
[\int_S(\overline{k}_{\alpha}{^m}{_{\beta}}+
\overline{k}'_{\alpha}{^m}{_{\beta}})(\phi^{\alpha}
-\phi'^{\alpha})(\phi^{\beta}-\phi'^{\beta})
\epsilon_{mabc} dS^{abc}]^{1/2} 
\nonumber\\ 
&&\ +2\sigma\tau||(\overline{k}-\overline{k}')
\nabla(\phi+\phi')e^{-t/\tau}||+4\sigma\tau
||(\overline{j}-\overline{j}')e^{-t/\tau}||\nonumber\\
&&\ +4\sigma\tau||(\overline{k}\nabla\phi+\overline{j})
e^{-t/\tau}|| +4\sigma\tau||(\overline{k}'\nabla\phi'
+\overline{j}')e^{-t/\tau}||,
\label{B.3}\end{eqnarray}
where we have suppressed indices.  The first term on the
right in (\ref{B.3}) is a surface term arising from
the integration by parts (the surface term at $S_{t_o}$
having been absorbed into the inequality).  In (\ref{B.3}), $||\ ||$ means ``the square root of the
integral of the square of the indicated field over $V$"
 (i.e., the L${^2}$-norm),
and $\tau$ and $\sigma$ are positive
constants such that
\begin{equation}
\tau\nabla_m(\overline{k}_{\alpha}{^m}{_{\beta}}
+\overline{k}'_{\alpha}{^m}{_{\beta}})\leq 
(\nabla_mt)(\overline{k}_{\alpha}{^m}{_{\beta}}+
\overline{k}'_{\alpha}{^m}{_{\beta}})\geq
\frac{1}{\sigma} G_{\alpha\beta},
\label{B.4}\end{equation}
everywhere in $V$. (Note that such constants exist, by
compactness of $V$ and positive-definiteness of the
middle expression in (\ref{B.4}).) The inequality (\ref{B.3})
is our final result.  It asserts
that two fields, $\phi$ and $\phi'$, are close to each
other in $V$ (left side of  (\ref{B.3}))
 provided that they are close on the initial
surface $S$ (first term on the right), that each approximately satisfies
an equation of the form 
(\ref{B.1}) (fourth and fifth terms on the right),
 and that their respective coefficients,
($\overline{k},\; \overline{j}$) and ($\overline{k}',
\; \overline{j}'$), in this equation are close 
(second and third terms on the right). It is in this
derivation of the inequality (\ref{B.3}) that we make
crucial use of the symmetric, positive-definite character
of the coefficients $k_{\alpha}{^m}{_{\beta}}$, and the
geometrical conditions of part c of the theorem.    

We first prove uniqueness (parts b and c of the theorem).
Let $\phi$ and $\phi'$ be two solutions, as in part b, and
let $\hat{U}\subset U\cap U'$ and the family $S_t$ be as
in part c.  Apply inequality (\ref{B.3}), with
$\overline{k}_{\alpha}{^m}{_{\beta}}=k_{\alpha}{^m}{_{\beta}}
(x,\phi(x))$, $\overline{k}'_{\alpha}{^m}
{_{\beta}}=k_{\alpha}{^m}{_{\beta}}(x,\phi'(x))$, 
$\overline{j}_{\alpha} = j_{\alpha}(x,\phi(x))$,
and $\overline{j}'_{\alpha}=j_{\alpha}(x,\phi'(x))$. 
 Then the first term
on the right vanishes (by initial conditions), and the last
two terms on the right vanish (by (\ref{B.1})).  But, for the
remaining two terms, each
of $|k(\phi)-k(\phi')|$ and $|j(\phi)-j(\phi')|$ is bounded by
a multiple (namely, the least upper bound of
$|\partial k/\partial\phi|$ and $|\partial j/\partial\phi |$,
respectively) of $|\phi-\phi'|$.  So, choosing $\tau$
sufficiently small, the sum of these two remaining terms
on the right of (\ref{B.3}) is less than the left side
of this inequality.  We thus conclude that $||(\phi-\phi')e^{-t/\tau}||
=0$, and so that $\phi=\phi'$.

We next turn to existence.  This is carried out in two steps.  

Consider first the equation          
\begin{equation}
\overline{k}_{\alpha}{^m}{_{\beta}}
\nabla_m\phi^{\beta}+\overline{j}_{\alpha}=0,
\label{B.5}\end{equation}
with fixed fields $\bar{k}_{\alpha}{^m}{_{\beta}}(x)$,
$\bar{j}_{\alpha}(x)$.  This is just (\ref{B.1}), but with the coefficients
evaluated on a fixed background field. 
Fix a family $S_t$ of surfaces, and a region $V$,
as in the derivation of Eqn. (\ref{B.3}).  Fix also a
field $S\stackrel{\phi_0}{\rightarrow}F$ on $S$,
and a positive number $\epsilon$.   We
wish to show existence of a smooth $\phi$ in $V$,
with $\phi|_S=\phi_0$, that is an ``$\epsilon$-approximate"
solution of (\ref{B.5}), i.e., that is such that the
square root of the integral over $V$ of the square of the left side is 
less than or equal to $\epsilon$.  We may set
$\phi_0=0$ (by replacing $\phi$ by $\phi+\hat{\phi}$ in (\ref{B.5}),
where $\hat{\phi}|_S=\phi_0$, and then absorbing into $\bar{j}_{\alpha}$ the 
extra term, $\overline{k}_{\alpha}{^m}{_{\beta}}
\nabla_m\hat{\phi}^{\beta}$, thus created). 
 Then, to show existence of an 
$\epsilon$-approximate solution of (\ref{B.5}), it suffices to
show that fields of the form $\overline{k}_{\alpha}{^m}{_{\beta}}
\nabla_m\phi{^{\beta}}$, with $\phi$ smooth
and vanishing on $S$, are L$^2$-dense in $V$.  But for this, in
turn, it suffices to show: Given any square-integrable
field $\psi^{\alpha}$ on $V$, such that
\begin{equation}
\int_V\psi^{\alpha}\overline{k}_{\alpha}{^m}{_{\beta}}
\nabla_m\phi^{\beta}\; = \; 0
\label{B.6}\end{equation}
for every smooth $\phi$ vanishing on $S$, we must have
$\psi^{\alpha}=0$. So, fix such a $\psi^{\alpha}$.  Were
this $\psi$ smooth, then we could proceed as follows.  Integrate
the left side of Eqn. (\ref{B.6}) by parts.  First
choosing $\phi$ to have support in the interior of $V$,
we obtain (from the volume integral) that 
$\nabla_m(\overline{k}_{\alpha}{^m}{_{\beta}}
\psi^{\beta})=0$; and then choosing $\phi$ to be nonzero
on $S_{t_o}$, we obtain (from the surface term) that
$\psi^{\alpha}|_{S_{t_o}}=0$.  But these two together imply
that $\psi^{\alpha}=0$.  Indeed, setting $\overline{k}' 
_{\alpha}{^m}{_{\beta}} = \overline{k}_{\alpha}{^m}{_{\beta}}$ 
, $\overline{j}_{\alpha}
= (\nabla_m\overline{k}_{\alpha}{^m}{_{\beta}})\psi^{\beta}$,
$\overline{j}'_{\alpha}= 0$, $\phi=\psi$, $\phi'=0$ 
in (\ref{B.3}), we obtain
\begin{eqnarray} 
||\psi e^{-t/\tau}||&\leq&\sqrt{\sigma\tau}\int_{S_{t_o}}
\overline{k}_{\alpha}{^m}{_{\beta}}\psi^{\alpha}
\psi^{\beta}\epsilon_{mabac}\; \; dS^{abc}\nonumber\\ 
&&+4\sigma\tau||(\nabla\overline{k})\psi
e^{-t/\tau}|| + 4\sigma\tau||\nabla
(\overline{k}\psi)e^{-t/\tau}||,
\label{B.7}\end{eqnarray}
while, for $\tau$ sufficiently small, the second term
on the right is less than or equal to the left side.  But, unfortunately, $\psi^{\alpha}$ need not be smooth,
but only square-integrable.  We therefore proceed as follows.
Let $h$ be a nonnegative, smooth, symmetric, two-point function
on $M$, such that $h(x,y)$ vanishes for $x$ and $y$ sufficiently
separated, and, for each fixed $y$, the integral of $h(x,y)$
over M has value one.  Set $\hat{\psi}(x)=\int_{M_y}h(x,y)\psi(y)$,
where the integration variable is $y\in M$.  This is a smooth
approximation to $\psi$.  Now take a sequence of such $h$'s
such that ``$h$, together with its derivative, approaches a 
delta-function and its derivative", in the sense that 
\begin{equation}
\int_{M_y}h(x,y)f(y)\; \rightarrow f(x),\ \ \nabla
\int_{M_y}h(x,y)f(y)\; \rightarrow\nabla f(x),
\label{B.8}\end{equation}
for every smooth function $f$ on $M$. Then the left side of
(\ref{B.6}), with $\psi$ replaced by $\hat{\psi}$, approaches
zero, and so, it can be shown, each of the first and third terms on the right in (\ref{B.7}),
with $\psi$ replaced by $\hat{\psi}$, approaches zero.
It now follows from (\ref{B.7}) (again, choosing $\tau$
sufficiently small) that $||\hat{\psi}e^{-t/\tau}||$ approaches zero, and so that
$\psi^{\alpha}=0$.  Thus, we have shown existence of an
$\epsilon$-approximate solution of Eqn. (\ref{B.5}), in
a certain region $V$, with given initial data.

We now adopt an iterative procedure.  Begin with any smooth
field $\phi_1$ on $V$, satisfying the initial condition,
$\phi_1|_S=\phi_0$.  Choose $\epsilon_2>0$, set
$\overline{k}_{\alpha}{^m}{_{\beta}} = k_{\alpha}
{^m}{_{\beta}}(x,\phi_1(x))$, $\bar{j}_{\alpha}=
j_{\alpha}(x,\phi_1(x))$ in (\ref{B.5}), and find an
$\epsilon_2$-approximate solution, $\phi_2$, of that
equation satisfying the initial condition.  Then choose
$\epsilon_3>0$, set $\overline{k} = k(\phi_2)$,
$\bar{j}=j(\phi_2)$ in (\ref{B.5}),
and find an $\epsilon_3$-approximate solution,
$\phi_3$, of that equation satisfying the initial condition.
Continuing in this way, with $\epsilon_i\rightarrow0$, we obtain
a sequence of fields, $\phi_1, \phi_2,\cdots$.  Each of
these satisfies the initial condition, and each approximately
satisfies (\ref{B.5}) (better, as $i$ increases) with
its predecessor as background.  We wish to show that the
$\phi_i$ converge to the desired solution.
To this end, set $\phi= \phi_i$, $\phi'=\phi_{i-1}$, 
$\overline{k}=k(\phi_{i-1})$, $\overline{k}'=k(\phi_{i-2})$,
$\overline{j}=j(\phi_{i-1})$, $\overline{j}'=j(\phi_{i-2})$
in (\ref{B.3}),
to obtain
\begin{eqnarray}
||\phi_i-\phi_{i-1}||&\leq&\sigma\tau e^{t_o/\tau}
[2||(k(\phi_{i-1})-k(\phi_{i-2}))\nabla
(\phi_i+\phi_{i-1})||\nonumber\\      
&& +4||j(\phi_{i-1})-j(\phi_{i-2})||
+4(\epsilon_i +\epsilon_{i-1})]\nonumber\\ 
&\leq&\sigma\tau e^{t_o/\tau}[\alpha
||\phi_{i-1}-\phi_{i-2}|| +4(\epsilon_i+\epsilon_{i-1})], 
\label{B.9}\end{eqnarray}
where we used $e^{-t_o/\tau}\leq e^{-t/\tau}\leq 1$ in the first
step; and set $\alpha = 2\; {\rm lub}|\frac{\partial k}{\partial \phi}|
{\rm lub}$ $|\nabla(\phi_i+\phi_{i-1})|+4\; {\rm lub}|\frac{\partial j}{\partial
\phi}|$ in the second.  Suppose for a moment that $\phi_i$
and $\nabla\phi_i$ were uniformly bounded (i.e., there is a single 
constant that bounds all the $\phi_i$ throughout $V$,
and similarly for $\nabla\phi_i$).  Then $\sigma$ (via
(\ref{B.4})) and $\alpha$ (above) would remain bounded as $i$
increases, and so, by choosing $\tau$, $t_o$, and the
$\epsilon_i$ sufficiently small\footnote{It is here that, by 
having to choose $t_o$ sufficiently small, we
restrict the size of the neighborhood $U$ for part (a) of the theorem. (In the
large-$t$ region of $V$, the $\phi_i$ could fail to converge,
indicating that there the final solution would become singular.)
This is the starting point for deriving a
strengthening
of part a guaranteeing a minimum size to its neighborhood $U$.} in 
(\ref{B.9}), we would guarantee convergence of 
$\sum||\phi_i-\phi_{i-1}||$, and so convergence of
the $\phi_i$ to some field $\phi$ on $V$. 
 But, unfortunately, we cannot, at this
stage, guarantee uniform boundedness of even the $\phi_i$
---much less of their derivatives.  Indeed, all we control
is a certain average of the $\phi_i$, represented by the
left side of (\ref{B.9}).  We therefore proceed as follows.  Taking
the $x$-derivative of Eqn. (\ref{B.1}), and introducing a new
field $\phi_a{^{\alpha}}(x)$ to represent ``$\nabla\phi$",
we obtain
\begin{equation}
k_{\alpha}{^m}{_{\beta}}\nabla_m(\phi_a{^{\beta}})+
\phi_m{^{\beta}}(\nabla_ak_{\alpha}{^m}{_{\beta}}
+\phi_a{^{\gamma}}\frac{\partial}{\partial\phi^{\gamma}}
k_{\alpha}{^m}{_{\beta}})\nonumber\\
+\nabla_aj_{\alpha} + \phi_a{^{\gamma}}
\frac{\partial}{\phi^{\gamma}}j_{\alpha}=0,
\label{B.10}\end{equation}
The combination of (\ref{B.1}) and (\ref{B.10}) is a quasilinear,
first-order system\footnote{For initial data on $S$, we take,
for $\phi^{\alpha}$, the given $\phi_0$, and, for
$\phi_a{^{\alpha}}$, ``$\nabla_a\phi^{\alpha}$", as
computed from $\phi_0$  and Eqn. (\ref{B.1})
evaluated on $S$.}  of partial differential equations for the 
fields $\phi^{\alpha}$, $\phi_a{^{\alpha}}$.  Furthermore,
it inherits from (\ref{B.1}) its hyperbolicity (since the
coefficient of the derivative-term in (\ref{B.10}) is the same
$k$ as in (\ref{B.1})).  Now continue in this way, taking
successive derivatives of (\ref{B.1}), introducing successive
fields, $\phi_{a\cdots c}{^{\alpha}}$, to represent
the higher derivatives of $\phi^{\alpha}$, and obtaining
successively larger hyperbolic systems.  Consider the system
that results after taking the fourth derivative---so 
the fields are now 
$\phi^{\alpha}$,$\phi_a{^{\alpha}}$, $\phi_{ab}{^{\alpha}}$,
$\phi_{abc}{^{\alpha}}$, $\phi_{abcd}{^{\alpha}}$ and
the equations (\ref{B.1}) and its first four derivatives---and
apply to this entire system the iterative procedure
above.  Then the left side of (\ref{B.9}) will include
a term $||\phi_{i\; abcd}{^{\alpha}}-\phi_{i-1\; abcd}{^{\alpha}}||$.
We now apply a Sobolev inequality\footnote{R. A. Adams, {\em Sobolev
Spaces} (Academic Press, NY, 1975).}, which asserts that
$||\nabla\nabla\nabla\nabla\phi||$ provides a uniform bound
on $\phi$ and its first two\footnote{We need, at this point,
a uniform bound on the second derivative of $\phi$ because
there appears in our equations, by this point, a term
``$(\nabla\nabla\phi)(\nabla\nabla\nabla\phi)$", and we have
on $\nabla\nabla\nabla\phi$ only an L${^2}$ bound.} derivatives.
From this, combined with (\ref{B.9}), it can be shown that
$\phi_i{^{\alpha}}$, $\phi_{i\; a}{^{\alpha}}$, 
$\phi_{i\; ab}{^{\alpha}}$, $\phi_{i\; abc}{^{\alpha}}$,
$\phi_{i\; abcd}{^{\alpha}}$ all converge, to some fields
$\phi^{\alpha}$, $\phi_a{^{\alpha}}$, $\phi_{ab}{^{\alpha}}$,
$\phi_{abc}{^{\alpha}}$, $\phi_{abcd}{^{\alpha}}$.  The
$\phi^{\alpha}$ that results from this procedure is the
desired solution of (\ref{B.1}).  To show that this 
$\phi^{\alpha}$ is smooth, take still higher derivatives
of (\ref{B.1}), and proceed
as above, introducing as new
fields still higher derivatives of $\phi$, and applying
the interative procedure above to the resulting hyperbolic
system. One must check that, in the demonstration of
existence for this succession of hyperbolic systems, the
number $t_o$ (which governs the size of the region $V$) can
remain bounded away from zero.   There results convergence of
$\phi_{i\; a\cdots c}{^{\alpha}}=\nabla_a\cdots \nabla_c
\phi_i{^{\alpha}}$, and so smoothness of $\phi^{\alpha}
= \lim\; \; \phi_i{^{\alpha}}$.    
\\

\noindent{\em Acknowledgement:}  I wish to thank Vivek Iyer for
numerous discussions on these topics.

\end{document}